\documentclass[pre,twocolumn,superscriptaddress,tightenlines,showpacs,longbibliography,floatfix,footinbib]{revtex4-1}

\usepackage{latexsym,amsmath,amsfonts,tabularx,amssymb,bm,empheq}

\usepackage{color}
\usepackage[dvipsnames]{xcolor}
\definecolor{nblue}{RGB}{28,130,185}
\definecolor{cgreen}{RGB}{76,153,0}
\definecolor{myorange}{RGB}{245,156,74}

\usepackage{hyperref}
\hypersetup{
  colorlinks=true,
  citecolor=magenta,
  urlcolor=-myorange
}

\usepackage{tensor}
\usepackage{ulem}
\usepackage{blkarray}
\usepackage{mathtools}
\usepackage{multirow}

\begin{document}
\title{Cell divisions imprint long lasting elastic strain fields in epithelial tissues}

\author{Ali Tahaei}
\affiliation{Max Planck Institute for the Physics of Complex Systems, 	N\"{o}thnitzer Str.38, 01187 Dresden, Germany}

\author{Romina Piscitello-G\'omez}
\affiliation{Max Planck Institute for Molecular Cell Biology and Genetics, Pfotenhauerstrasse 108, Dresden, 01307, Germany}

\author{S Suganthan}
\affiliation{Max Planck Institute for the Physics of Complex Systems, 	N\"{o}thnitzer Str.38, 01187 Dresden, Germany}

\author{Greta Cwikla}
\affiliation{Excellence Cluster, Physics of Life, Technische Universit\"{a}t Dresden, Arnoldstrasse 18, Dresden, 01307, Germany}

\author{Jana F. Fuhrmann}
\affiliation{Max Planck Institute for Molecular Cell Biology and Genetics, Pfotenhauerstrasse 108, Dresden, 01307, Germany}
\affiliation{Excellence Cluster, Physics of Life, Technische Universit\"{a}t Dresden, Arnoldstrasse 18, Dresden, 01307, Germany}

\author{Natalie A. Dye}
\affiliation{Max Planck Institute for Molecular Cell Biology and Genetics, Pfotenhauerstrasse 108, Dresden, 01307, Germany}
\affiliation{Excellence Cluster, Physics of Life, Technische Universit\"{a}t Dresden, Arnoldstrasse 18, Dresden, 01307, Germany}

\author{Marko Popovi\'c}
\affiliation{Max Planck Institute for the Physics of Complex Systems, 	N\"{o}thnitzer Str.38, 01187 Dresden, Germany}
\affiliation{Excellence Cluster, Physics of Life, Technische Universit\"{a}t Dresden, Arnoldstrasse 18, Dresden, 01307, Germany}
\affiliation{Center for Systems Biology Dresden, Pfotenhauerstrasse 108, 01307 Dresden, Germany}

\begin{abstract}
A hallmark of biological tissues, viewed as complex cellular materials, is the active generation of mechanical stresses by cellular processes, such as cell divisions. Each cellular event generates a force dipole that deforms the surrounding tissue. Therefore, a quantitative description of these force dipoles, and their consequences on tissue mechanics, is one of the central problems in understanding the overall tissue mechanics. In this work, we analyze previously published experimental data on fruit fly \textit{D. melanogaster} wing epithelia to quantitatively describe the deformation fields induced by a cell-scale force dipole. We find that the measured deformation field can be explained by a simple model of wing epithelium as a linearly elastic sheet. This allows us to infer the magnitude and dynamics of the mechanical forces generated by the cell divisions. In particular, we find that cell divisions exert a transient isotropic force dipole field, corresponding to the temporary localization of the cell nucleus to the tissue surface during the division, and traceless-symmetric force dipole field that remains detectable from the tissue strain field for up to about $3.5$ hours after the division. This is the timescale on which elastic strains are erased by other mechanical processes and therefore it corresponds to the tissue fluidization timescale. In summary, we have developed a method to infer force dipoles induced by cell divisions, by observing the strain field in the surrounding tissues. Using this method we quantitatively characterize mechanical forces generated during a cell division and their effects on the tissue mechanics.
\end{abstract}

\maketitle

\section{Introduction}
How mechanical forces influence biological tissues is one of the central problems in animal development and regeneration \cite{Heisenberg2013, Brugues2014, LeGoff2016, Julicher2017, Tetley2019}. Biological tissues are often described as soft viscoelastic materials \cite{Aigouy2010} that are elastic on sufficiently short timescales. Tissue fluidization on longer timescales arises from cell divisions, extrusions, and intercalations, which restructure the tissue over time \cite{Ranft2010, Etournay2015, Curran2017,Kim_2021}. 

Cellular processes, such as cell divisions and cell extrusions, are intrinsic sources of mechanical force generation within tissues, inducing deformation in the surrounding tissue \cite{Gibson2011,Mao2013,Roellig2022}, which can in turn lead to cellular rearrangements \cite{Firmino2016} and contribute to the overall tissue flow \cite{Ranft2010,Matoz-Fernandez2017,BocanegraMoreno2023}. Furthermore, cell divisions have been indicated to control the glassy cell dynamics observed in cultured epithelial tissues \cite{Angelini2011,Park2015} and explanted embryonic tissues \cite{Schotz2013}. However, despite the fact that cell divisions are a hallmark of biological matter and seem to be a major factor in controlling material properties of biological tissues, a quantitative understanding of the mechanical forces generated by cell divisions and their immediate consequences on the surrounding tissue is still lacking.  

To study the mechanics of cell divisions, we need to probe mechanical stresses in the surrounding tissue. One of the most common methods of probing mechanical stresses is the laser ablation experiment, in which one or multiple cells are destroyed by a laser. The resulting displacement of the ablated tissue boundary reflects the pre-stresses existing in the tissue before the ablation \cite{Farhadifar2007,mayer_anisotropies_2010,Bonnet2012,Kasza2014,Etournay2015,Piscitello2022}. Pre-stresses are then characterised either by the initial dynamics of tissue boundary displacement \cite{mayer_anisotropies_2010,Kasza2014} or from the final relaxed shape of the tissue \cite{Farhadifar2007,Bonnet2012,Dye2021,Piscitello2022}. Previous analyses focused on changes in the position or shape of specific points or outlines in the vicinity of the ablation. Therefore, they could not provide an insight into the strain field induced in the surrounding tissue, and they had to rely on assumptions about tissue mechanical properties. Furthermore, to extract information about all components of the pre-stress in the tissue, previous methods require a precise, circular laser ablation \cite{Bonnet2012,Dye2021}. If we could instead infer force dipoles induced by cellular processes such as cell divisions, we would be able to extract information about tissue mechanics in a completely non-invasive way. Therefore, the starting point of our work, on the way to study cell division mechanics, is to develop a method that allows us to infer the force dipole in an epithelial tissue from the tissue strain field.  The idea to infer properties of a localized force field from the strain field it induces was recently employed to characterize the properties of the core of plastic events in computer glasses \cite{Moriel2020,Chacko2021,Moriel2024}. Here, we proceed independently of these results using an approach better suited for epithelial tissue data.

In this work, we first provide a detailed quantitative analysis of the tissue displacement field following a linear laser ablation performed in the pupal wing epithelium of the fruit fly \textit{D. melanogaster}. We show that the response of the fly wing epithelium to a laser ablation is consistent with that of a two-dimensional linear elastic sheet. This allows us to infer the force dipole induced by the ablation, normalized by the elastic constant, from the observed strain field. Furthermore, by analyzing the dynamics of the strain field, we show that the dominant mode of dissipation in the fly wing epithelium on short timescales less than a minute is viscous dissipation, as opposed to frictional dissipation with a substrate. 

Using our method to infer force dipole from its strain field, we determine the force dipole generated by cell divisions in the \textit{D. melanogaster} wing disc. We find that the isotropic component of the cell division force dipole tensor is only transient and vanishes almost immediately after the division, indicating that no net cell growth is accumulated during the cell division. Furthermore, we find that the traceless-symmetric component of the force dipole, imprinted by the cell division, remains visible in the surrounding tissue for up to about $3.5$ hours after the division. 
This provides a measurement of the fluidization timescale in the tissue and demonstrates the importance of elastic interactions in tissue-scale morphogenetic processes that take place on a comparable timescale, such as wing disc eversion \cite{Fuhrmann2023}.

\section{The fruit fly wing epithelium is a 2D linear elastic sheet}
\subsection{Linear laser ablation experiments}

\begin{figure*}
    \centering
    \includegraphics[width=\linewidth]{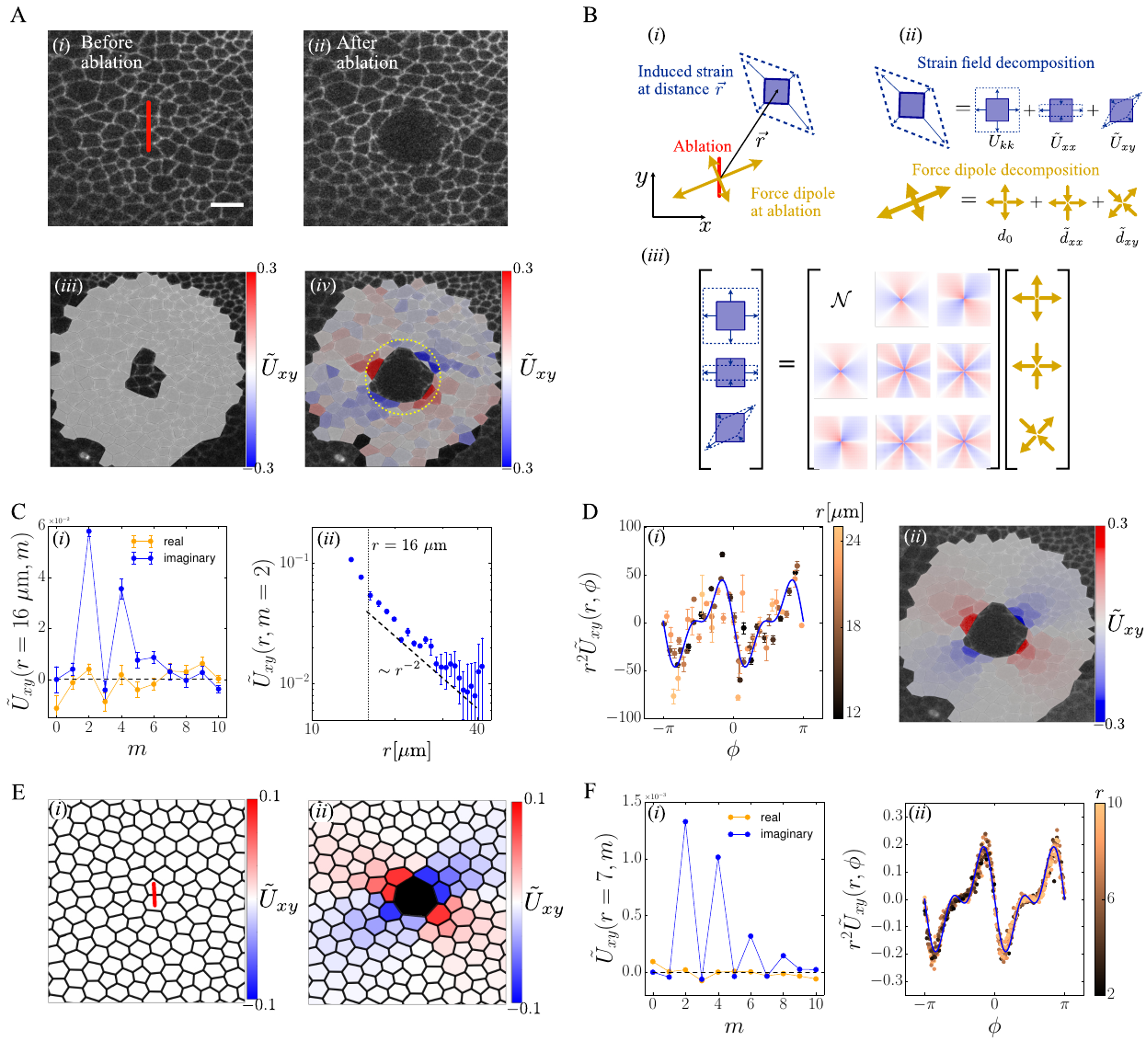}
    \caption{(A.\textit{i}) Fluorescently labeled cell outlines show the cellular packing before the laser ablation. The red bar indicates the location that will be ablated. (A.\textit{ii}) Cell outlines $50 s$ after the laser ablation. (A.\textit{iii}) Tracked and segmented cells in the region of interest are colored in white. (A.\textit{iv}) The accumulated cell strain component $U_{xy}$, $50 s$ after the laser ablation. The scale bar is $10 \mu m$.
    (B.\textit{i}) A force dipole in a two-dimensional linear elastic system induces a strain in distance $\vec r$. (B.\textit{ii}) Decomposition of two-dimensional strain field and force dipole tensors into isotropic and traceless-symmetric components. (B.\textit{iii}) A Voigt representation of the relation between strain at position $\vec{r}$ induced by a force dipole at the origin.
    (C.\textit{i}) Angular Fourier modes of $\tilde U_{xy}$ at distance $r= 16 ~\rm{\mu m}$, which corresponds to the yellow circle A.\textit{iv}, show a clear signal in modes $m=2$ and $m=4$. (C.\textit{ii}) The radial profile of the $m=2$ angular mode shows a radial profile consistent with $r^{-2}$ (dashed line). The dotted line indicates the distance $r=16~\rm{\mu m}$. Analysis in both C.\textit{i} and C.\textit{ii} is performed on the strain field averaged over the time interval from $30s$ to $70s$ after laser ablation and the error bars show the standard deviation over this interval. 
    (D.\textit{i}) The fit of the linear elastic theory to the experimental data, shown in C. The solid blue line is the fitted strain field with the corresponding fit parameters reported in the text. (D.\textit{ii}) Visualization of the fitted strain field. 
    (E.\textit{i}) The cellular network in the vertex model simulation. The red bar indicates two cells removed by ablation. (E.\textit{ii}) The cell strain field accumulated in the tissue after reaching the steady state following the simulated ablation.
    (F.\text{i}) Angular Fourier modes of the strain field pattern in E.\textit{ii} showing peaks in the second and fourth modes. 
    (F.\text{ii}) The fit of the linear elastic theory to the cell strain field in the vertex model. The solid blue is the fitted strain field with the fit parameters $d_0= 0.72 \pm 0.07  a_0 $, $\tilde{d}_{xx}= 0.65 \pm 0.07 a_0 $, $\tilde{d}_{xy}= 0.00 \pm 0.07 a_0 $ and $\mu = 1.3 \pm 0.1 $. The uncertainty intervals for fit parameters are obtained by bootstrapping the data (see Methods).}
    \label{fig:ablation}
\end{figure*}

In linear laser ablation experiments, a small linear segment of tissue is destroyed by a laser. In Ref. \cite{Piscitello2022}, laser ablation experiments  have  been performed in the \textit{D. melanogaster} wing epithelium during pupal morphogenesis. Information about the pre-stresses in the tissue was inferred from the dynamics of the nearest cell-cell junction as it retracted away from the ablation. Here, we use laser ablation experiment data published in Ref. \cite{Piscitello2022} to study the deformation field in the tissue surrounding the laser ablation. 
An image of a wing tissue before and after such a laser ablation experiment is shown in Fig. \ref{fig:ablation} (A.\textit{i}), where the red line indicates the line ablated by the laser. 
Ablations were performed in a plane thinner than $1~\mu m$ on the apical surface of the tissue \cite{Piscitello2022}. Following the ablation, the tissue relaxes until reaching a mechanical equilibrium after about $T_e \approx 30s$ after ablation. We show an example of the relaxed state in Fig. \ref{fig:ablation} (A.\textit{ii}). The tissue remains in this state until after around $1 \rm{min}$. After this time, the tissue starts to exhibit additional flows, directed towards the line of ablation, which likely correspond to the wound-healing response of the tissue, see Methods and SI. Note that in most experiments from Ref. \cite{Piscitello2022} we could not segment and analyze a reasonably large circle of cells around the ablation (SI Fig. 1), we selected four experiments that we could best analyze. In the main text we present analysis of the clearest one, denoted Experiment 1. In Experiment 2, such flows appear earlier so that no intermediate relaxed state can be identified. In Experiments 3 and 4 the strain field generated by the ablation is also weaker than in Experiment 1, so that the noise overcomes the signal already at small distances from the ablation. Analyses of these three experiments are presented in the SI.

To measure the tissue strain field, we segment the apical cell surfaces over time and quantify area $a$ and the elongation tensor $Q_{ij}$ of each cell. The cell elongation tensor is defined such that the co-rotational changes in the elongation tensor $Q_{ij}$ correspond to the accumulated pure shear strain \cite{Etournay2015,Merkel2017}.
Since no cell divisions or rearrangements appear in the region of interest following the laser ablation, we quantify the tissue strain field ${U}_{ij}(\bm{r}, t)$ from the change in cell shape and size relative to their values before the ablation, see Methods. The strain field tensor consists of the isotropic strain component $U_{kk}$ and traceless-symmetric component $\tilde{U}_{ij}$ which we denote as pure shear strain component. In Fig. \ref{fig:ablation} (A.\textit{iv}) we show the pure shear strain field component $\tilde{U}_{xy}$ at $50s$ after the laser ablation, which exhibits a clear spatial pattern. See SI Fig. 3 (A) for the other traceless-symmetric component and the isotropic component.

Our goal is to find the relationships between the induced strain field and the stresses induced by the ablation. Since the process of ablation does not introduce any net force, the dominant component of the induced force field will be dipolar. Therefore, we aim to relate components of the strain field at distance $\vec{r}$ from the ablation to the components of the force dipole tensor, as illustrated in Fig.~\ref{fig:ablation} (B.\textit{i,ii}). 

We first consider tissue strain after it has reached the mechanical equilibrium after the time $T_e$ and calculate the discrete angular Fourier spectrum $\tilde{U}^{ss}_{ij}(r,m)$, where $r = |\bm{r}|$ is the distance from the center of ablation and $m$ is the angular mode index, see Methods.
We find the strongest signal in the second and the fourth mode of the pure shear strain components, see Fig.~\ref{fig:ablation} (C.\textit{i}) and SI Fig.~3 (B.\textit{ii}), which correspond to $\cos(2\varphi)$ and $\cos(4\varphi)$ angular Fourier modes. In the isotropic strain component, we find the strongest signal in the zeroth and second mode, see SI Fig.~3 (B.\textit{i}), which correspond to the constant and $\cos(2\varphi)$ angular Fourier modes. 
We then quantify the radial dependence of the second angular mode and we find that its radial profile is consistent with $r^{-2}$, as shown in Fig.~\ref{fig:ablation} (C.\textit{ii}). These results reminded us of the far field of Eshelby propagators in two-dimensional linear elastic sheets \cite{Eshelby1957,Nicolas2018,Moriel2020}, and we next tested whether the linear elastic theory can indeed quantitatively account for our measurements.

We briefly present the linear elastic theory of a sheet in which a point force dipole has been inserted. A force dipole is represented by a tensor $D_{ij}$ that can be decomposed into isotropic and traceless-symmetric contributions $D_{ij}= D_0\delta_{ij} + \tilde{D}_{ij}$ \footnote{We consider only force dipoles that do not exert a net torque.}. Inserting a force dipole in an elastic sheet induces a strain field that is determined by the force balance equation
\begin{align}   \partial_j\sigma_{ij}+\partial_j(\delta(\bm{r})D_{ij})&= 0
\end{align}
where $\sigma_{ij}$ is the stress tensor in the elastic sheet, which can be decomposed into pressure $P$ and shear stress $\tilde{\sigma}$: $\sigma_{ij}= -P\delta_{ij} + \tilde{\sigma}_{ij}$. The resulting strain field can be written as
\begin{align}\label{eq:2dElasticity}
    u_{ij}(\bm{r})= G_{ijkl}(\bm{r}) D_{kl},
\end{align}
where $G_{ijkl}$ is the elastic dipole propagator and $\bm{r}$ is the distance from the force dipole, see SI. We graphically illustrate the components of Eq. \ref{eq:2dElasticity} in Fig. \ref{fig:ablation} (B.\textit{iii}), using the Voigt tensor representation. In this representation, the force dipole propagator elements are organized in a matrix, which we plot in Fig. \ref{fig:ablation} (B). From this point, we use the normalized force dipole defined as $d_{ij}=D_{ij}/2K$. The propagator elements depend only on $\cos(2\varphi)$, $\sin(2\varphi)$, $\cos(4 \varphi)$ and $\sin(4\varphi)$, see Fig. \ref{fig:ablation} (B.\textit{iii}) and SI, consistent with the observed second and fourth angular Fourier modes in the data.
Furthermore, all components of $G_{ijkl}$ decay in space as $r^{-2}$. This is consistent with the the radial profile observed in the experimental data for $m=2$ modes, see Fig.~\ref{fig:ablation} (C.\textit{ii}) and SI Fig.~5. The radial profiles of $m=4$ modes are more noisy so that it is difficult to characterise the radial decay.

Therefore, the strain field measured in the laser ablation experiments is consistent with a linear elastic response of the tissue to the force dipoles generated by the ablation. However, to confirm this hypothesis, we need to show that linear elastic theory can quantitatively account for the observed strain fields. 

We fit the steady state tissue strain field $U^{ss}_{ij}$ with a strain field predicted by a linear elastic theory of a two-dimensional elastic sheet in Eq. \ref{eq:2dElasticity}. The fitting parameters are the three independent components of the force dipole tensor and the ratio $\mu$ of the two-dimensional shear and bulk elastic moduli. By analyzing the ablation experiment, we find that the data are well described by the fit, as shown in Fig.~\ref{fig:ablation} (D) and SI Fig. 6 (A). The fitted parameter values are: $d_0= 160\pm 10 ~\mu m^2$, $\tilde{d}_{xx}= 150 \pm 10 ~\mu m^2$, $\tilde{d}_{xy}= 20 \pm 10 ~\mu m^2$, $\mu= 1.9\pm 0.1$, see Methods. 
Furthermore, the ratio $\mu \approx 2$, corresponding to the Poisson ratio of $\nu \approx -0.33$, is comparable with the value $\mu \approx 3$, corresponding to the Poisson ratio of $\nu \approx -0.5$, recently reported in the same tissue using a laser ablation method based on circular ablation shape \cite{Piscitello2022}. 

To further test our analysis of the laser ablation experiments, we simulate a linear ablation in the vertex model of epithelial tissues \cite{Farhadifar2007}. We remove two cells from a pre-stressed cellular network, where the pre-stress represents the proximal-distal stress in the wing epithelium \cite{Etournay2015}, and the ablation axis is perpendicular to the pre-stress orientation, recapitulating the orientation of ablation in experiments, see SI for details. We then quantify the resulting strain and find that the simulation shows the same pattern as in experiments, consistent with the linear elastic theory of a 2D sheet Fig.~\ref{fig:ablation} (E, F) and SI Fig. 7.  We fit the linear elastic theory to the observed strain field pattern and infer the force density generated by the ablation from tissue deformation. Finally, we find the Poisson's ratio of the vertex model cellular network to be $\nu = -0.13 \pm 0.04$. This finding is consistent with the value $\nu = -0.095 \pm 0.009$ found by a direct numerical measurement of the shear and bulk modulus of the same network, see SI.

Our results show that the response of the \textit{D. melanogaster} wing epithelium to a small linear laser ablation is consistent with that of a linearly elastic sheet with a negative two-dimensional Poisson ratio. This allows us to infer the force dipole tensor induced by the ablation from the tissue strain field, which we confirmed using the vertex model simulations.

\subsection{Viscous dissipation determines tissue relaxation dynamics}
Our analysis of the strain in the ablated tissue after it has relaxed allows us to study its elastic properties. Now, we explore the transient dynamics of tissue strain as it approaches the mechanical equilibrium, which will allow us to study dissipative processes in fly wing epithelium. We consider two possible dissipation mechanisms: viscosity and friction with a substrate. 

\begin{figure}[ht!]
    \centering
    \includegraphics[width=\linewidth]{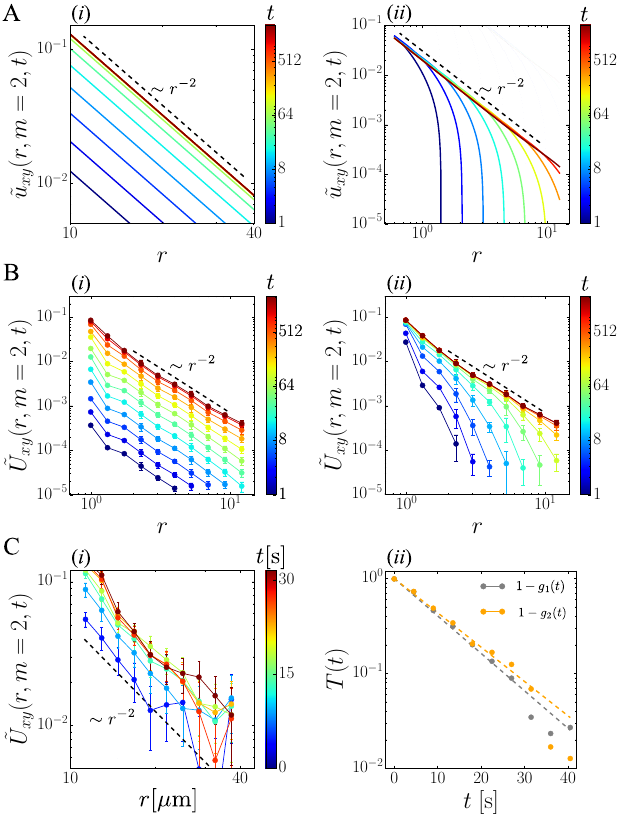}
    \caption{Time evolution the second Fourier mode $m=2$ of the strain field $xy$ component induced by a force dipole for: (A.\textit{i}) linear elastic sheet with viscous dissipation, (A.\textit{ii}) linear elastic sheet with frictional dissipation, vertex model simulation with
    (B.\textit{i}) viscous dissipation and, (B.\textit{ii}) frictional dissipation. (C) Laser ablation experiment. (C.i) Time evolution of $\tilde{U}_{xy}$ shows a radial profile as $r^{-2}$, consistent with a viscous dissipation mechanism. (C.ii) Quantifying the viscous relaxation of strain as described in the text, which allows the measurement of viscous relaxation timescales. The black dashed line indicates the power-law $r^{-2}$ decay for reference. Each data point represents the strain field average over $5$ smaller bins and the errors are the corresponding standard deviations over those bins.}
    \label{fig:ablation_relaxation}
\end{figure}

Interestingly, in a continuum model of tissue as an elastic sheet, these two dissipation mechanisms lead to qualitatively different relaxation dynamics upon insertion of a force dipole. In particular, in an elastic sheet with only viscous dissipation, all strain field components decay as $1/r^2$ with the distance from the force dipole throughout the relaxation with the amplitude evolving in time, see SI and Fig. \ref{fig:ablation_relaxation} (A.\textit{i}). However, in an elastic sheet with only frictional dissipation, the strain field propagates through the sheet: at distances within the propagation front, the strain field converges to the steady state solution and rapidly decays beyond the front, see SI and Fig. \ref{fig:ablation_relaxation}
(A.\textit{ii}).

To compare the strain field dynamics in viscous and frictional dissipation models within a cellular system, we performed vertex model simulations of laser ablation that include only the frictional dissipation and compare it with a viscous dissipation, see SI for details. As discussed above, the signature difference between viscous and frictional dissipation is the power law decay of the shape $1/r^2$ in the former and the existence of a propagation front in the latter.
With a frictional dissipation, we indeed find a clear propagation of the strain field that converges to the steady state value as the propagation front expands (\ref{fig:ablation_relaxation} (B.\textit{ii}) and SI Fig. 12 (B)), consistent with the analytical calculation (\ref{fig:ablation_relaxation} (A.\textit{ii})).
In contrast, with viscous dissipation, we observe that the strain field decays as $1/r^2$ in all time points (\ref{fig:ablation_relaxation} (B.\textit{i}) and SI Fig. 12 (A)).

To quantify the dynamics of tissue relaxation following the laser ablation, we measure the strain field in time intervals of $4.5 s$, starting from $1 s$ after the ablation, (SI Fig.~8). We find that the angular strain pattern is the same as in the mechanical equilibrium one, see Movie 1. 
We analyze the dynamics of the strain radial profile by calculating the second angular Fourier mode of the strain field at all measured timepoints, see Fig.~\ref{fig:ablation_relaxation} (C.\textit{i}) and SI Fig.~8. We find that the strain field is consistent with $1/r^2$ decay and we do not see any signature of a propagating front, see Fig. 2. Therefore, we conclude that the viscosity is the dominant mode of dissipation in the fly wing epithelium on timescales below $T_e \approx 30s$.

Our continuum model of an elastic sheet provides the dynamical strain propagator ${G}_{ij}(\bm{r}, t)={G}_{ij}(\bm{r}) {T}_{ij}(t)$, where no sum over repeated indices is performed. ${G}_{ij}(\bm{r})$ is the steady state propagator, and ${T}_{ij}(t)$ is the relaxation matrix whose components contain two relaxation factors $g_1(t)= 1 - \exp{(-t/\tau_1)}$ and $g_2(t)=1 - \exp{(-t/\tau_2)}$, where the two relaxation timescales 
\begin{align}
    \tau_1&= \frac{\overline{\eta}}{\overline{K}}\frac{1 + 2\eta/\overline{\eta}}{1 + \mu}\\
    \tau_2&= \frac{\eta}{K}
\end{align}
are related to the bulk and shear elastic constants, and bulk and shear viscosities $\overline{K}, 2K, \overline{\eta}, \text{ and } 2\eta$, respectively (see SI I.B.(ii)).
We extract $\tau_1$ and $\tau_2$ from experimental data, see Methods, SI, and SI Fig.~13.
We find that the two relaxation timescales are very similar: $\tau_1= 11\pm 1 s$ and $\tau_2= 12\pm 1 s$, where intervals of uncertainty correspond to the uncertainty of the fit, see Methods. 

As mentioned above, the Experiment 1 presented here was selected based on our ability to segment and track cells in a radius of $40~\mu m$ surrounding the laser ablation, which was not possible for most of the other experiments. Here, we highlight that in Experiment 2, the angular and radial pattern of the strain field are consistent with that of an elastic sheet with viscous dissipation on times comparable to the relaxation timescales $\tau_1$ and $\tau_2$ extracted above, as reported in the SI Fig. 13. Interestingly, on longer times radial profile of strain departs from $r^{-2}$ and continues to evolve throughout the experiment. Similar additional motion is observed even in the Experiment 1 reported above, but only after about $2~\rm{min}$, see SI Fig. 11. This suggests that additional dynamics are due to a wound-healing response in the tissue and understanding this behavior would require a dedicated experimental exploration in the future.

In summary, our detailed analysis of the strain in the fly wing epithelium in response to the laser ablation shows that the tissue behaves as a two-dimensional viscoelastic solid. These results suggest that we can now infer force dipoles generated by cell divisions, and thereby study their mechanics, by simply observing the dynamics of the surrounding tissue strain caused by the division, without the need to directly perturb the tissue.

\section{Elastic deformation imprinted by a cell division}
Cell division is a complex process during which the mother cell is split into two daughter cells. 
Here, we aim to characterize the mechanical forces generated by the dividing cell from the analysis of the strain field that the dividing cell induces in the epithelium. 
We analyze cell divisions in previously published experiments \cite{Dye2021} on explanted \textit{D. melanogaster} wing imaginal discs.  
We are interested in the generic features of the response of epithelial tissues to a mechanical perturbation, such as the laser ablation or a cell division. Therefore, we aim to extract normalised force dipoles in the surface plane generated by cell divisions using the same approach as with the laser ablation experiments. However, we perform the analysis independently of any results obtained from laser ablations in the pupal wing.

The wing disc developmental stage precedes the pupal stage in which linear laser ablations analysed above were performed \cite{Piscitello2022}, and their mechanical moduli may be different. 
However, a quantification of the strain field generated by cell divisions is more convenient in the wing disc stage compared to the pupal wing stage.
Cell divisions in the pupal wing differ from divisions in the wing disc in that they are patterned in time and space, with most cells dividing at least once within several hours and the tissue exhibits strong tissue shear flows and active cell rearrangements \cite{Etournay2016}. Therefore, analysis of the strain field induced by individual divisions is practically impossible in the pupal wing due to additional dynamics in the tissue.
However, in both tissues, the apical surfaces are mechanically special as they contain a high concentration of actomyosin cortex and adherens junctions between cells that are important for morphogenesis. Therefore, it is likely that apical mechanics plays a significant role in the overall tissue mechanics in both tissues.

\begin{figure}
    \centering
    \includegraphics[width=\linewidth]{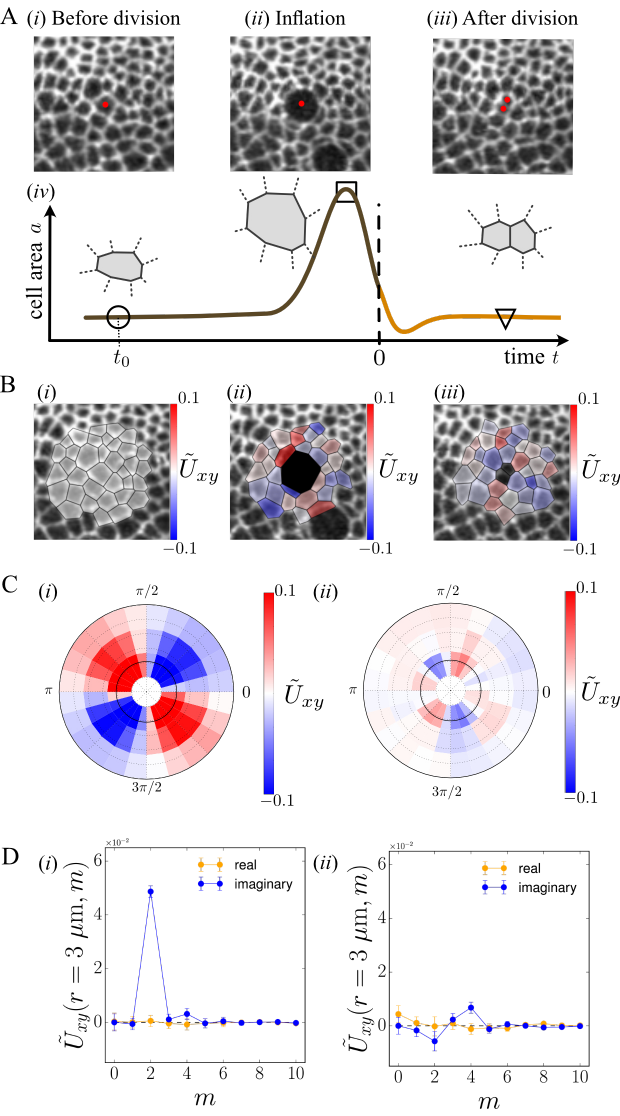}
    \caption{(A) Snapshots of a cell division in three stages. (A.\textit{i}) As the mother cell starts the division process, (A.\textit{ii}) it rounds up, corresponding to the inflation, and then (A.\textit{iii}) it shrinks and divides into the daughter cells. The red dots show the mother and the daughter cells. (A.\textit{iv}) A schematic representation of the cell area of the mother cell (brown) and the daughter cells (orange) during the division process. The symbols highlight the stage of the dividing cell before the division, at the peak of inflation, and after the division. The new bond forms at $t=0$, and $t_0$ is the reference state. (B) Strain $\tilde{U}_{xy}$ of the cells near a typical cell division at the three stages shown in (A) with $t_0=-60 ~ [\rm{min}]$. (C) Ensemble average ($N=682$) of $\tilde{U}_{xy}$ around the dividing cell: (C.\textit{i}) at the peak of the inflation $t=-20 ~ [\rm{min}]$ (corresponding to (A.\textit{ii}) and (B.\textit{ii})), and (C.\textit{ii}) after the division $t=30 ~ [\rm{min}]$ (corresponding to (A.\textit{iii}) and (B.\textit{iii})). The radial extent of the plots is $6 \mu m$. (D) The angular Fourier transform of the $\tilde{U}_{xy}$ at a distance of $r=3 ~ \mu m$ from the center of the mother cell at $t=-20 ~ [\rm{min}]$ (D.\textit{i}) and at $t=30 ~  [\rm{min}]$ (D.\textit{ii}).}
    \label{fig:division_ExpData}
\end{figure}

Wing discs were cultured over about $13$ hours, during which the tissue was imaged every $5$ minutes, and individual cells were segmented and tracked over time. 
The progress of a typical cell division is shown in Fig.~\ref{fig:division_ExpData} (A). Before the actual cell division, the mother appears to inflate Fig.~\ref{fig:division_ExpData} (A.ii). This is not a real increase in cell size, but is rather a consequence of the cell nucleus arriving at the apical surface of the cell to divide and transiently stretching the cell apical surface \cite{Gibson2011, Jeon2016}. This transient inflation is followed by the creation of the new cell boundary, and the two daughter cells become visible Fig.~\ref{fig:division_ExpData} (A.iii). For each cell division, we define the time of the new bond creation as the reference cell division time $t= 0$. 

Fig.~\ref{fig:division_ExpData} (A.\textit{iv}) qualitatively summarizes the experimental observations. 
For each cell division, we quantify the tissue strain field in the vicinity of the dividing cell, as described in Section II, relative to the starting timepoint $t_0=-60 ~\rm{min}$ before the cell division time. We show an example of the cell strain field induced by a dividing cell in Fig. \ref{fig:division_ExpData} (B). The timescale on which the cell division forces evolve, which is on the order of tens of minutes, is much larger than the minutes relaxation timescale reported in the wing disc \cite{Dye2021}. Therefore, the strain field dynamics observed over tens of minutes is largely due to changes of the force dipole generated by the cell division, and not due to much faster stress propagation through the tissue. Although it is possible that on these longer timescales cells adapt to the experienced stress and adapt their reference shape, a clear positive correlation between stress and strain has been reported in our past work \cite{Dye2021}. Therefore, it is reasonable to interpret the change in cell elongation as a measure of elastic strain, even if the adaptation to stress effectively renormalizes the elastic constants, as discussed in Ref. \cite{Dye2021}.


\begin{figure}
    \centering
    \includegraphics[width=\linewidth]{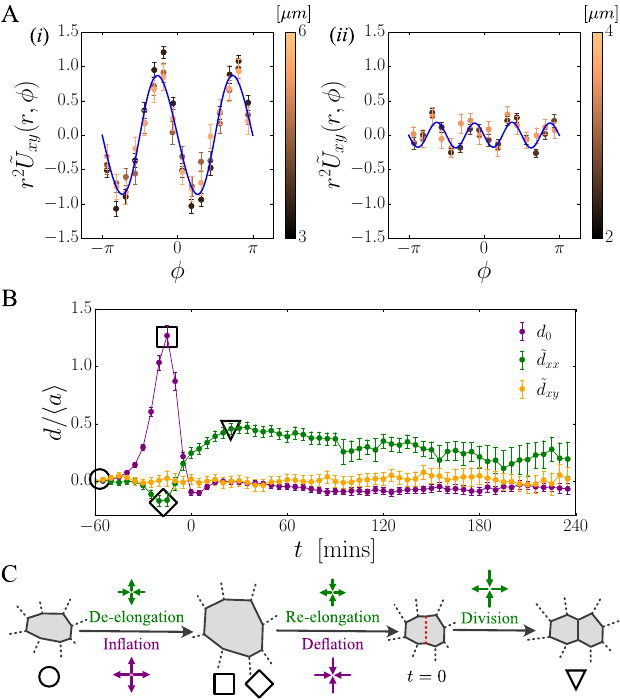}
    \caption{(A) The fit of linear elastic theory to the measured strain field around cell divisions at (A.\textit{i}) the peak of inflation $t=-20 ~ [\rm{min}]$ and (A.\textit{ii}) after division $t=30 ~ [\rm{min}]$. Each data point shows the average strain field in polar bins and the error bars are the standard deviation over these bins. (B) The force dipole generated by a cell division as a function of time. (C) A schematic summary of the force dipoles during stages of a cell division.}
    \label{fig:division_TheoryData}
\end{figure}

We find that a single cell division generates weaker forces compared to those induced by laser ablation (compare Fig.~\ref{fig:ablation} A and Fig.~\ref{fig:division_ExpData} B), and the corresponding strain field is more susceptible to noise.
Therefore, we average the strain field from multiple cell divisions ($N=682$). We first align the centers of the cell division, defined as the area-weighted average of daughter cell centers at the cell division time ($t=0$). We next rotate each division such that the orientation of the line connecting daughter cell centers is oriented along the $x-$axis at the cell division time.
We show the resulting strain field component $\tilde{U}_{xy}$ in Fig.~\ref{fig:division_ExpData} (C) and the corresponding angular Fourier modes (D) at times $t= -20~\rm{min}$, corresponding to the peak of the cell inflation, and $t= 30 ~\rm{min}$ after the cell division.
In Fig.~\ref{fig:division_ExpData} (C.\textit{i}), the 2-fold symmetry of the angular strain pattern indicates that the isotropic force dipole is dominant, consistent with the isotropic cell inflation. In Fig.~\ref{fig:division_ExpData} (C.\textit{ii}), the pattern is closer to a 4-fold symmetric one, such as would be induced by an anisotropic force dipole. Finally, we find that radial profiles of the Fourier modes are consistent with $r^{-2}$ decay (SI Fig. 15), as was the case with laser ablation experiments above.

We fit the linear elastic theory Eq.~\ref{eq:2dElasticity} to the experimentally measured strain field at each time point and extract the resulting force dipole $d_{ij}(t)$. Comparison between the fit and the data is shown for the $xy$ component of the strain field at $t=-20 \rm{min}$ and $t=30 \rm{min}$ in Fig. \ref{fig:division_TheoryData} B.
We find that the isotropic force dipole component $d_0$ rises sharply during the mother cell inflation and peaks at about $t= -20 ~\rm{min}$ (magenta line in Fig. \ref{fig:division_TheoryData} B). However, by the time of the new bond formation, the isotropic component has disappeared and remains negligible at later times. This result shows that cells do not noticeably grow during the division, but rather that the division reorganizes the available material into the two daughter cells.
The traceless part of the force dipole $\tilde{d}_{xx}$ grows as the new bond is formed and persists in the tissue multiple hours after the division (green line in Fig. \ref{fig:division_TheoryData} B), indicating that the division-induced strains remain imprinted in the tissue long after the division has been executed. 
Over time, $\tilde{d}_{xx}$ slowly decays while its fluctuations increase, representing the effect of other cell divisions and rearrangements that appear in the tissue. We estimate the tissue fluidization time from the decay of $\tilde{d}_{xx}$ in time by fitting an exponential function to it. We find the fluidization timescale to be $\tau_\alpha \simeq 3.5 \pm 0.7$ hours (see Methods), over which $\tilde{d}_{xx}$ vanishes and, therefore, the strain field induced by the division is erased. Note that component $\tilde{d}_{xy}$ remains negligible at all times. 

Finally, we note that at the time of the peak of $d_0$, there is a small but clear negative peak in $\tilde{d}_{xx}$. To interpret this observation, we have to remember that our strain measurements are relative to the state of the cells at $t_0= -60 ~\rm{min}$. Since the mother cell attains an almost perfectly circular shape during the inflation, any initial anisotropy of the mother cell at time $t_0$ gives rise to an anisotropic force dipole component during the inflation. Therefore, the observed negative value of $\tilde{d}_{xx}$ indicates that mother cells are elongated along the future division axis, consistent with the well-known Hertwig's rule \cite{Hertwig1884,Thery2007,Gibson2011,Lisica2022,Middelkoop2023}. We summarise these results in Fig. \ref{fig:division_TheoryData} (C).

\section{Discussion}

In this work, we devised a full spatio-temporal analysis of tissue strain induced by two types of force dipole fields: linear laser ablations and cell divisions. 
In this way, we have shown that the developing \textit{Drosophila} wing epithelium can be described as a two-dimensional elastic solid with viscous dissipation. We also measured its two-dimensional Poisson ratio, which we find to be negative, consistent with the values reported in the literature \cite{Dye2021,Piscitello2022}.
Our analysis was based on laser ablation of the wing tissue during pupal development, at which time the wing consists of two apposed epithelial layers. Although the experimental observations are consistent with a model of the wing apical surface as a linear elastic sheet, in the future it might be interesting to probe the three-dimensional structure response of the tissue with deeper ablations and in wings where the two epithelial layers have been separated \cite{Sun2021}.

We characterized the dynamics of mechanical forces generated by a dividing cell throughout the cell division process. We found that the strain fields imprinted by a cell division are slowly erased on the timescale of $\tau_\alpha= 3.5\pm 0.7 ~\rm{hr}$, which can be understood as a fluidization timescale of the \textit{Drosophila} wing tissue. Furthermore, our analysis shows that there is no cell growth during a division, and stresses induced by the cell division correspond to those of a pure shear force dipole. Finally, we found that Hertwig's rule applies in the \textit{D. melanogaster} wing disc \cite{Gibson2011, Mao2013}.

An important implication of our work is that the \textit{Drosophila} wing epithelium is a solid whose fluidization time $\tau_\alpha$ is not negligible on the developmental timescales. For example, large-scale shape changes that the wing undergoes during eversion are executed on a similar timescale \cite{Fuhrmann2023}. A possible physical picture of the \textit{Drosophila} wing epithelium that could account for our observations is that of a glassy material, or as was put forward in the glass literature, a solid that flows \cite{Dyre2023,Ozawa2023,Tahaei2023}. In other words, the long-range elastic interactions between cells would lead to correlated glassy dynamics, which has been reported in tissue cultures \cite{Angelini2011,Park2015}, and it remains to be seen whether they indeed play a role in developing tissues. In summary, our results quantify force dipoles generated by dividing cells in developing \textit{D. melanogaster} wing epithelia, as well as elastic interactions between cells, which paves the way to the development of tissue mechanics theory and motivates further experimental study of collective cell dynamics.

\section*{Acknowledgements}
We would like to thank Michael Staddon, Frank J\"ulicher, Jonas Neipel for useful discussions and comments. Funding for the biological experiments was supported by the Max Planck Society and the Deutsche Forschungsgemeinschaft (Cluster of Excellence Physics of Life, EXC-2068-390729961 and SPP1782/EA4/10-2). NAD additionally acknowledges funding from Deutsche Krebshilfe/MSNZ Dresden. MP and ND acknowledge funding by the Deutsche Forschungsgemeinschaft (DFG, German Research Foundation) - Project number DY 180/2-1 (eBer-24-23102) and PO 3023/2-1 - 544201605.

\section{Methods}

\subsection{Cell segmentation and tracking}

The linear ablation experiments were performed during the pupal morphogenesis stage and were previously published in Ref. \cite{Piscitello2022}, where the authors analyzed the initial retraction velocity of the ablated bonds. Here, we re-analyzed movies of those experiments, and fully segmented and tracked cells in the entire field using the FIJI plugin, TissueAnalyzer \cite{Aigouy2016}, using a bond cutoff of 2 pixels. Then, after the manual correction of segmentation errors, we use TissueMiner \cite{Etournay2016} to analyze changes in the cell shape and size.

The cell division experimental data came from the wing disc stage and was previously segmented, tracked, and analyzed in Ref. \cite{Dye2017,Dye2021}.

\subsection{Tissue relaxation after a laser ablation}
After the ablation, nearby tissue relaxes to a new mechanical equilibrium. In Experiment 1 reported in the main text, we estimate equilibrium to be reached after about $30s$ after ablation, when the tissue does not visibly evolve any more. However, this equilibrium state is not permanent, as at about $1$ minute after the ablation, we observe additional motion of the tissue, which we interpret as the onset of the wound-healing response. In this work, we are interested in the stress response to the ablation, not the wound-healing process, and therefore we limit our analysis to times below $1$ minute after the laser ablation. Note that in Experiments 2-4, reported in detail in SI, the additional patterns appear to emerge earlier and we could not extract as clear an intermediate regime as in Experiment 1. However, even in these experiments the profiles of strain observed during the relaxation are consistent with the linear elastic theory, thereby further strengthening our results from Experiment 1.

\subsection{Tissue strain field measurement}
Strain field consists of isotropic strain $U_{kk}(\mathbf{r}, t)/2$ and shear strain $\tilde{U}_{ij}(\mathbf{r}, t)$.
To determine the strain field from the changes of cell shape and area, we use the method based on the cell triangulation \cite{Etournay2015, Merkel2017, Duclut2021} in which the co-rotational change of the cell elongation tensor $Q_{ij}$ corresponds to the tissue shear rate, and changes in the logarithm of cell area to the isotropic strain rate. Here, we approximate the accumulated tissue strain by the change in cell elongation tensor and in the logarithm of cell area for pure shear and isotropic components, respectively. To define the cell elongation tensor $Q_{ij}$, we choose the sub-cellular triangulation of the tissue obtained by connecting two adjacent vertices to the geometric center of cells, as described in \cite{Duclut2021}.
We estimate the accumulated tissue shear strain field in a cell from the difference between the cell shape tensors 
\begin{align}
    \tilde{U}_{ij}(t)&\simeq Q_{ij}(t) - Q_{ij}(0)
\end{align}
The only approximation is that we neglect to account for changes in cell elongation due to local rotations. We determined the exact cell shear tensor at several timepoints, and we found that the differences from our approximation are small. Therefore, we use the approximate expression for simplicity.
The isotropic strain field in a cell is defined as the accumulated relative area change rate \cite{Etournay2015,Merkel2017}, which we measure relative to the initial cell state as
\begin{align}
    U_{kk}(t)&= \log{\left(\frac{a(t)}{a(0)}\right)} \quad ,
\end{align}
where $a$ is the cell area.

Finally, we define the tissue strain field $U_{ij}(\mathbf{r}, t)$ at position $\mathbf{r}$ to have the value of the strain tensor obtained for the cell inside of which point $\mathbf{r}$ lies.

\subsection{Discrete angular Fourier transform}
We examine the angular discrete Fourier transformation of the tissue strain field to investigate its angular pattern and study its radial profiles. 
The strain in position $\bm{r}$ is represented by ${U}_{ij}(r, \varphi)$, where $r$ is the distance to the center of the cut and $\varphi$ is the polar angle.
We use the notation ${U}_{ij}(r, m)$ for the $m$-th angular mode of discrete angular Fourier transform at a distance $r$, and it is defined as,
\begin{align}
    {U}_{ij}(r,m) = \frac{1}{N} \sum_{i=1}^{N} {U}_{ij}(r, \varphi_i) e^{im\varphi_i},
\end{align}
where $\varphi_i$ are $N$ uniformly distributed numbers between $0$ and $2\pi$. In rare cases a point with coordinates $(r, \varphi_i)$ does not belong to a cell, making ${U}_{ij}(r, \varphi)$ undefined at that point. To handle such cases, we modify the Fourier transform as follows,
\begin{align}
    {U}_{ij}(r,m) = \frac{\sum_{i=1}^{N} {U}_{ij}(r, \varphi_i) e^{im\varphi_i} \iota(r, \varphi_i)}{\sum_{i=1}^N 
    \iota(r, \varphi_i)},
\end{align}
where $\iota(r, \varphi_i)=1$ if the point belongs to a cell and $0$ otherwise.

For the analyses of cell division data, a point can belong to multiple cells due to simultaneous analysis of multiple cell divisions. In such cases, we replace $U_{ij}(r,\varphi_i)$ with its average $\langle U_{ij}(r,\varphi_i) \rangle$ where $\langle \cdots \rangle$ is average over all divisions.

\subsection{Fit of the elastic theory to data}\label{sec:fitting_process}
We measure the force dipole, the ratio of shear moduli to bulk moduli, and viscosity by fitting the strain field pattern to the elastic and viscoelastic theories. 
First, we define a cost function 
\begin{multline}
    E = \sum_{\alpha \in \rm{cells}} \Big[  (U_{kk}^\alpha - u_{kk}(\bm{r}^\alpha))^2  + ( \tilde{U}_{xx}^\alpha - \tilde{u}_{xx}(\bm{r}^\alpha) )^2 +\\
    (\tilde{U}_{xy}^\alpha - \tilde{u}_{xy}(\bm{r}^\alpha))^2 \Big],
    \label{eq:cost_func}
\end{multline}
where ${U}_{ij}^\alpha$ is the cell strain measured from data, $\bm{r}^\alpha$ is the position of the center of each cell and ${u}_{ij}(\bm{r}^\alpha)$ is the strain field predicted from theory at position $\bm{r}^\alpha$. 
By minimizing $E$, we find the magnitude of the force dipole, the ratio of the shear modulus to bulk modulus, and when fitting the relaxation dynamics, the relaxation timescales. 

When fitting the steady state strain field measured in  Experiment 1, we wanted to take into account the fluctuations in the steady state. For this, we sampled the strain field at 5 timepoints uniformly spaced between $36s$ and $58.5s$ after ablation. We constructed datasets by bootstrapping the sampled data $10$ times, and we fit the linear elastic theory for each dataset. The reported parameter values and uncertainties are mean values and standard deviations of these 10 fit outcomes.

To fit the relaxation dynamics and extract the relaxation timescale, we consider the strain field at each time point $t_i$, spaced $4.5 s$ apart, during the relaxation and use the steady state parameters and force dipole to fit the elements of the matrix $\boldsymbol{T}(t_i)$. Finally, we fit an exponential function to the obtained matrix element values at different time points, Fig. 2.C.\textit{ii}, and report the relaxation timescale values and their uncertainties obtained by this exponential fit.

To analyze cell divisions, we first reduce noise in the strain field by applying a moving average over a $10$-minute window. Next, we select the time point $t=30 \rm{min}$ and systematically vary the reference time $t_0$ from $-120 \rm{min}$ to $-40 \rm{min}$. For each $t_0$ within this range, we fit the steady state linear elastic theory to the measured strain field to extract the corresponding set of $\mu(t_0)$ values. The parameter $\mu$ is then determined as the average  $\langle\mu(t_{0})\rangle$, which remains fixed for the subsequent strain field analyses (see SI Fig. 14). This is done to more precisely measure the parameter $\mu$, which cannot be extracted independently of $d_0$ when traceless symmetric components of the force dipole vanish. Namely, in that case it is impossible to factorize the product $\mu D_0$ as the two factors only appear in the form of the product, see SI Eqs. 9 and 10.  Finally, to compute the force dipole at each timepoint $t$ we set $t_0=-60 \rm{min}$, and we perform the fit at each time point using the bootstrapping algorithm over the ensemble of all divisions, obtaining the force dipole values and their uncertainties at each time point, while using the previously obtained values of $\mu$ as a constant. 


\subsection{Estimation of the tissue fluidization time}
To measure the decay timescale of the anisotropic force dipole $\tilde{d}_{xx}$, we fit an exponential function to normalized force dipole $y=\tilde{d}_{xx}(t) / \langle a \rangle$ as $f(y) = \tilde{d}_{xx}(t=0)\exp[-t/\tau_\alpha]$ to observed data in SI Fig.~16. From that, we obtain the fluidization timescale $\tau_\alpha = 3.5 \pm 0.7 \rm{hr}$ and $\tilde{d}_{xx}(t=0)/\langle a \rangle = 0.53 \pm 0.01$. The reported errors correspond to the range of parameters over which an exponential function successfully fits the data.

\bibliography{bib.bib}

\begin{thebibliography}{46}%
\makeatletter
\providecommand \@ifxundefined [1]{%
 \@ifx{#1\undefined}
}%
\providecommand \@ifnum [1]{%
 \ifnum #1\expandafter \@firstoftwo
 \else \expandafter \@secondoftwo
 \fi
}%
\providecommand \@ifx [1]{%
 \ifx #1\expandafter \@firstoftwo
 \else \expandafter \@secondoftwo
 \fi
}%
\providecommand \natexlab [1]{#1}%
\providecommand \enquote  [1]{``#1''}%
\providecommand \bibnamefont  [1]{#1}%
\providecommand \bibfnamefont [1]{#1}%
\providecommand \citenamefont [1]{#1}%
\providecommand \href@noop [0]{\@secondoftwo}%
\providecommand \href [0]{\begingroup \@sanitize@url \@href}%
\providecommand \@href[1]{\@@startlink{#1}\@@href}%
\providecommand \@@href[1]{\endgroup#1\@@endlink}%
\providecommand \@sanitize@url [0]{\catcode `\\12\catcode `\$12\catcode `\&12\catcode `\#12\catcode `\^12\catcode `\_12\catcode `\%12\relax}%
\providecommand \@@startlink[1]{}%
\providecommand \@@endlink[0]{}%
\providecommand \url  [0]{\begingroup\@sanitize@url \@url }%
\providecommand \@url [1]{\endgroup\@href {#1}{\urlprefix }}%
\providecommand \urlprefix  [0]{URL }%
\providecommand \Eprint [0]{\href }%
\providecommand \doibase [0]{http://dx.doi.org/}%
\providecommand \selectlanguage [0]{\@gobble}%
\providecommand \bibinfo  [0]{\@secondoftwo}%
\providecommand \bibfield  [0]{\@secondoftwo}%
\providecommand \translation [1]{[#1]}%
\providecommand \BibitemOpen [0]{}%
\providecommand \bibitemStop [0]{}%
\providecommand \bibitemNoStop [0]{.\EOS\space}%
\providecommand \EOS [0]{\spacefactor3000\relax}%
\providecommand \BibitemShut  [1]{\csname bibitem#1\endcsname}%
\let\auto@bib@innerbib\@empty
\bibitem [{\citenamefont {Heisenberg}\ and\ \citenamefont {Bellaïche}(2013)}]{Heisenberg2013}%
  \BibitemOpen
  \bibfield  {author} {\bibinfo {author} {\bibfnamefont {Carl-Philipp}\ \bibnamefont {Heisenberg}}\ and\ \bibinfo {author} {\bibfnamefont {Yohanns}\ \bibnamefont {Bellaïche}},\ }\bibfield  {title} {\enquote {\bibinfo {title} {Forces in tissue morphogenesis and patterning},}\ }\href {\doibase 10.1016/j.cell.2013.05.008} {\bibfield  {journal} {\bibinfo  {journal} {Cell}\ }\textbf {\bibinfo {volume} {153}},\ \bibinfo {pages} {948–962} (\bibinfo {year} {2013})}\BibitemShut {NoStop}%
\bibitem [{\citenamefont {Brugués}\ \emph {et~al.}(2014)\citenamefont {Brugués}, \citenamefont {Anon}, \citenamefont {Conte}, \citenamefont {Veldhuis}, \citenamefont {Gupta}, \citenamefont {Colombelli}, \citenamefont {Muñoz}, \citenamefont {Brodland}, \citenamefont {Ladoux},\ and\ \citenamefont {Trepat}}]{Brugues2014}%
  \BibitemOpen
  \bibfield  {author} {\bibinfo {author} {\bibfnamefont {Agustí}\ \bibnamefont {Brugués}}, \bibinfo {author} {\bibfnamefont {Ester}\ \bibnamefont {Anon}}, \bibinfo {author} {\bibfnamefont {Vito}\ \bibnamefont {Conte}}, \bibinfo {author} {\bibfnamefont {Jim~H.}\ \bibnamefont {Veldhuis}}, \bibinfo {author} {\bibfnamefont {Mukund}\ \bibnamefont {Gupta}}, \bibinfo {author} {\bibfnamefont {Julien}\ \bibnamefont {Colombelli}}, \bibinfo {author} {\bibfnamefont {José~J.}\ \bibnamefont {Muñoz}}, \bibinfo {author} {\bibfnamefont {G.~Wayne}\ \bibnamefont {Brodland}}, \bibinfo {author} {\bibfnamefont {Benoit}\ \bibnamefont {Ladoux}}, \ and\ \bibinfo {author} {\bibfnamefont {Xavier}\ \bibnamefont {Trepat}},\ }\bibfield  {title} {\enquote {\bibinfo {title} {Forces driving epithelial wound healing},}\ }\href {\doibase 10.1038/nphys3040} {\bibfield  {journal} {\bibinfo  {journal} {Nature Physics}\ }\textbf {\bibinfo {volume} {10}},\ \bibinfo {pages} {683–690} (\bibinfo {year} {2014})}\BibitemShut {NoStop}%
\bibitem [{\citenamefont {LeGoff}\ and\ \citenamefont {Lecuit}(2016)}]{LeGoff2016}%
  \BibitemOpen
  \bibfield  {author} {\bibinfo {author} {\bibfnamefont {Loïc}\ \bibnamefont {LeGoff}}\ and\ \bibinfo {author} {\bibfnamefont {Thomas}\ \bibnamefont {Lecuit}},\ }\bibfield  {title} {\enquote {\bibinfo {title} {Mechanical forces and growth in animal tissues},}\ }\href {\doibase 10.1101/cshperspect.a019232} {\bibfield  {journal} {\bibinfo  {journal} {Cold Spring Harbor Perspectives in Biology}\ }\textbf {\bibinfo {volume} {8}},\ \bibinfo {pages} {a019232} (\bibinfo {year} {2016})}\BibitemShut {NoStop}%
\bibitem [{\citenamefont {Jülicher}\ and\ \citenamefont {Eaton}(2017)}]{Julicher2017}%
  \BibitemOpen
  \bibfield  {author} {\bibinfo {author} {\bibfnamefont {Frank}\ \bibnamefont {Jülicher}}\ and\ \bibinfo {author} {\bibfnamefont {Suzanne}\ \bibnamefont {Eaton}},\ }\bibfield  {title} {\enquote {\bibinfo {title} {Emergence of tissue shape changes from collective cell behaviours},}\ }\href {\doibase 10.1016/j.semcdb.2017.04.004} {\bibfield  {journal} {\bibinfo  {journal} {Seminars in Cell \& Developmental Biology}\ }\textbf {\bibinfo {volume} {67}},\ \bibinfo {pages} {103–112} (\bibinfo {year} {2017})}\BibitemShut {NoStop}%
\bibitem [{\citenamefont {Tetley}\ \emph {et~al.}(2019)\citenamefont {Tetley}, \citenamefont {Staddon}, \citenamefont {Heller}, \citenamefont {Hoppe}, \citenamefont {Banerjee},\ and\ \citenamefont {Mao}}]{Tetley2019}%
  \BibitemOpen
  \bibfield  {author} {\bibinfo {author} {\bibfnamefont {Robert~J.}\ \bibnamefont {Tetley}}, \bibinfo {author} {\bibfnamefont {Michael~F.}\ \bibnamefont {Staddon}}, \bibinfo {author} {\bibfnamefont {Davide}\ \bibnamefont {Heller}}, \bibinfo {author} {\bibfnamefont {Andreas}\ \bibnamefont {Hoppe}}, \bibinfo {author} {\bibfnamefont {Shiladitya}\ \bibnamefont {Banerjee}}, \ and\ \bibinfo {author} {\bibfnamefont {Yanlan}\ \bibnamefont {Mao}},\ }\bibfield  {title} {\enquote {\bibinfo {title} {Tissue fluidity promotes epithelial wound healing},}\ }\href {\doibase 10.1038/s41567-019-0618-1} {\bibfield  {journal} {\bibinfo  {journal} {Nature Physics}\ }\textbf {\bibinfo {volume} {15}},\ \bibinfo {pages} {1195--1203} (\bibinfo {year} {2019})}\BibitemShut {NoStop}%
\bibitem [{\citenamefont {Aigouy}\ \emph {et~al.}(2010)\citenamefont {Aigouy}, \citenamefont {Farhadifar}, \citenamefont {Staple}, \citenamefont {Sagner}, \citenamefont {Röper}, \citenamefont {Jülicher},\ and\ \citenamefont {Eaton}}]{Aigouy2010}%
  \BibitemOpen
  \bibfield  {author} {\bibinfo {author} {\bibfnamefont {Benoît}\ \bibnamefont {Aigouy}}, \bibinfo {author} {\bibfnamefont {Reza}\ \bibnamefont {Farhadifar}}, \bibinfo {author} {\bibfnamefont {Douglas~B.}\ \bibnamefont {Staple}}, \bibinfo {author} {\bibfnamefont {Andreas}\ \bibnamefont {Sagner}}, \bibinfo {author} {\bibfnamefont {Jens-Christian}\ \bibnamefont {Röper}}, \bibinfo {author} {\bibfnamefont {Frank}\ \bibnamefont {Jülicher}}, \ and\ \bibinfo {author} {\bibfnamefont {Suzanne}\ \bibnamefont {Eaton}},\ }\bibfield  {title} {\enquote {\bibinfo {title} {Cell {Flow} {Reorients} the {Axis} of {Planar} {Polarity} in the {Wing} {Epithelium} of {Drosophila}},}\ }\href {\doibase 10.1016/j.cell.2010.07.042} {\bibfield  {journal} {\bibinfo  {journal} {Cell}\ }\textbf {\bibinfo {volume} {142}},\ \bibinfo {pages} {773--786} (\bibinfo {year} {2010})}\BibitemShut {NoStop}%
\bibitem [{\citenamefont {Ranft}\ \emph {et~al.}(2010)\citenamefont {Ranft}, \citenamefont {Basan}, \citenamefont {Elgeti}, \citenamefont {Joanny}, \citenamefont {Prost},\ and\ \citenamefont {J{\" u}licher}}]{Ranft2010}%
  \BibitemOpen
  \bibfield  {author} {\bibinfo {author} {\bibfnamefont {J.}~\bibnamefont {Ranft}}, \bibinfo {author} {\bibfnamefont {M.}~\bibnamefont {Basan}}, \bibinfo {author} {\bibfnamefont {J.}~\bibnamefont {Elgeti}}, \bibinfo {author} {\bibfnamefont {J.-F.}\ \bibnamefont {Joanny}}, \bibinfo {author} {\bibfnamefont {J.}~\bibnamefont {Prost}}, \ and\ \bibinfo {author} {\bibfnamefont {F.}~\bibnamefont {J{\" u}licher}},\ }\bibfield  {title} {\enquote {\bibinfo {title} {Fluidization of tissues by cell division and apoptosis},}\ }\href {\doibase 10.1073/pnas.1011086107} {\bibfield  {journal} {\bibinfo  {journal} {Proceedings of the National Academy of Sciences}\ }\textbf {\bibinfo {volume} {107}},\ \bibinfo {pages} {20863–20868} (\bibinfo {year} {2010})}\BibitemShut {NoStop}%
\bibitem [{\citenamefont {Etournay}\ \emph {et~al.}(2015)\citenamefont {Etournay}, \citenamefont {Popovi{\' c}}, \citenamefont {Merkel}, \citenamefont {Nandi}, \citenamefont {Blasse}, \citenamefont {Aigouy}, \citenamefont {Brandl}, \citenamefont {Myers}, \citenamefont {Salbreux}, \citenamefont {J{\" u}licher},\ and\ \citenamefont {et~al.}}]{Etournay2015}%
  \BibitemOpen
  \bibfield  {author} {\bibinfo {author} {\bibfnamefont {Rapha{\" e}l}\ \bibnamefont {Etournay}}, \bibinfo {author} {\bibfnamefont {Marko}\ \bibnamefont {Popovi{\' c}}}, \bibinfo {author} {\bibfnamefont {Matthias}\ \bibnamefont {Merkel}}, \bibinfo {author} {\bibfnamefont {Amitabha}\ \bibnamefont {Nandi}}, \bibinfo {author} {\bibfnamefont {Corinna}\ \bibnamefont {Blasse}}, \bibinfo {author} {\bibfnamefont {Benoît}\ \bibnamefont {Aigouy}}, \bibinfo {author} {\bibfnamefont {Holger}\ \bibnamefont {Brandl}}, \bibinfo {author} {\bibfnamefont {Gene}\ \bibnamefont {Myers}}, \bibinfo {author} {\bibfnamefont {Guillaume}\ \bibnamefont {Salbreux}}, \bibinfo {author} {\bibfnamefont {Frank}\ \bibnamefont {J{\" u}licher}}, \ and\ \bibinfo {author} {\bibnamefont {et~al.}},\ }\bibfield  {title} {\enquote {\bibinfo {title} {Interplay of cell dynamics and epithelial tension during morphogenesis of the drosophila pupal wing},}\ }\href {\doibase 10.7554/eLife.07090} {\bibfield  {journal} {\bibinfo  {journal} {eLife}\ }\textbf {\bibinfo {volume} {4}} (\bibinfo {year} {2015}),\ 10.7554/eLife.07090}\BibitemShut {NoStop}%
\bibitem [{\citenamefont {Curran}\ \emph {et~al.}(2017)\citenamefont {Curran}, \citenamefont {Strandkvist}, \citenamefont {Bathmann}, \citenamefont {de~Gennes}, \citenamefont {Kabla}, \citenamefont {Salbreux},\ and\ \citenamefont {Baum}}]{Curran2017}%
  \BibitemOpen
  \bibfield  {author} {\bibinfo {author} {\bibfnamefont {Scott}\ \bibnamefont {Curran}}, \bibinfo {author} {\bibfnamefont {Charlotte}\ \bibnamefont {Strandkvist}}, \bibinfo {author} {\bibfnamefont {Jasper}\ \bibnamefont {Bathmann}}, \bibinfo {author} {\bibfnamefont {Marc}\ \bibnamefont {de~Gennes}}, \bibinfo {author} {\bibfnamefont {Alexandre}\ \bibnamefont {Kabla}}, \bibinfo {author} {\bibfnamefont {Guillaume}\ \bibnamefont {Salbreux}}, \ and\ \bibinfo {author} {\bibfnamefont {Buzz}\ \bibnamefont {Baum}},\ }\bibfield  {title} {\enquote {\bibinfo {title} {Myosin ii controls junction fluctuations to guide epithelial tissue ordering},}\ }\href {\doibase 10.1016/j.devcel.2017.09.018} {\bibfield  {journal} {\bibinfo  {journal} {Developmental Cell}\ }\textbf {\bibinfo {volume} {43}},\ \bibinfo {pages} {480--492.e6} (\bibinfo {year} {2017})},\ \bibinfo {note} {publisher: Elsevier}\BibitemShut {NoStop}%
\bibitem [{\citenamefont {Kim}\ \emph {et~al.}(2021)\citenamefont {Kim}, \citenamefont {Pochitaloff}, \citenamefont {Stooke-Vaughan},\ and\ \citenamefont {Campàs}}]{Kim_2021}%
  \BibitemOpen
  \bibfield  {author} {\bibinfo {author} {\bibfnamefont {Sangwoo}\ \bibnamefont {Kim}}, \bibinfo {author} {\bibfnamefont {Marie}\ \bibnamefont {Pochitaloff}}, \bibinfo {author} {\bibfnamefont {Georgina~A.}\ \bibnamefont {Stooke-Vaughan}}, \ and\ \bibinfo {author} {\bibfnamefont {Otger}\ \bibnamefont {Campàs}},\ }\bibfield  {title} {\enquote {\bibinfo {title} {Embryonic tissues as active foams},}\ }\href {\doibase 10.1038/s41567-021-01215-1} {\bibfield  {journal} {\bibinfo  {journal} {Nature Physics}\ }\textbf {\bibinfo {volume} {17}},\ \bibinfo {pages} {859--866} (\bibinfo {year} {2021})}\BibitemShut {NoStop}%
\bibitem [{\citenamefont {Gibson}\ \emph {et~al.}(2011)\citenamefont {Gibson}, \citenamefont {Veldhuis}, \citenamefont {Rubinstein}, \citenamefont {Cartwright}, \citenamefont {Perrimon}, \citenamefont {Brodland}, \citenamefont {Nagpal},\ and\ \citenamefont {Gibson}}]{Gibson2011}%
  \BibitemOpen
  \bibfield  {author} {\bibinfo {author} {\bibfnamefont {William~T.}\ \bibnamefont {Gibson}}, \bibinfo {author} {\bibfnamefont {James~H.}\ \bibnamefont {Veldhuis}}, \bibinfo {author} {\bibfnamefont {Boris}\ \bibnamefont {Rubinstein}}, \bibinfo {author} {\bibfnamefont {Heather~N.}\ \bibnamefont {Cartwright}}, \bibinfo {author} {\bibfnamefont {Norbert}\ \bibnamefont {Perrimon}}, \bibinfo {author} {\bibfnamefont {G.~Wayne}\ \bibnamefont {Brodland}}, \bibinfo {author} {\bibfnamefont {Radhika}\ \bibnamefont {Nagpal}}, \ and\ \bibinfo {author} {\bibfnamefont {Matthew~C.}\ \bibnamefont {Gibson}},\ }\bibfield  {title} {\enquote {\bibinfo {title} {Control of the mitotic cleavage plane by local epithelial topology},}\ }\href {\doibase 10.1016/j.cell.2010.12.035} {\bibfield  {journal} {\bibinfo  {journal} {Cell}\ }\textbf {\bibinfo {volume} {144}},\ \bibinfo {pages} {427--438} (\bibinfo {year} {2011})}\BibitemShut {NoStop}%
\bibitem [{\citenamefont {Mao}\ \emph {et~al.}(2013)\citenamefont {Mao}, \citenamefont {Tournier}, \citenamefont {Hoppe}, \citenamefont {Kester}, \citenamefont {Thompson},\ and\ \citenamefont {Tapon}}]{Mao2013}%
  \BibitemOpen
  \bibfield  {author} {\bibinfo {author} {\bibfnamefont {Yanlan}\ \bibnamefont {Mao}}, \bibinfo {author} {\bibfnamefont {Alexander~L}\ \bibnamefont {Tournier}}, \bibinfo {author} {\bibfnamefont {Andreas}\ \bibnamefont {Hoppe}}, \bibinfo {author} {\bibfnamefont {Lennart}\ \bibnamefont {Kester}}, \bibinfo {author} {\bibfnamefont {Barry~J}\ \bibnamefont {Thompson}}, \ and\ \bibinfo {author} {\bibfnamefont {Nicolas}\ \bibnamefont {Tapon}},\ }\bibfield  {title} {\enquote {\bibinfo {title} {Differential proliferation rates generate patterns of mechanical tension that orient tissue growth},}\ }\href {\doibase https://doi.org/10.1038/emboj.2013.197} {\bibfield  {journal} {\bibinfo  {journal} {The EMBO Journal}\ }\textbf {\bibinfo {volume} {32}},\ \bibinfo {pages} {2790--2803} (\bibinfo {year} {2013})}\BibitemShut {NoStop}%
\bibitem [{\citenamefont {Roellig}\ \emph {et~al.}(2022)\citenamefont {Roellig}, \citenamefont {Theis}, \citenamefont {Proag}, \citenamefont {Allio}, \citenamefont {Bénazéraf}, \citenamefont {Gros},\ and\ \citenamefont {Suzanne}}]{Roellig2022}%
  \BibitemOpen
  \bibfield  {author} {\bibinfo {author} {\bibfnamefont {Daniela}\ \bibnamefont {Roellig}}, \bibinfo {author} {\bibfnamefont {Sophie}\ \bibnamefont {Theis}}, \bibinfo {author} {\bibfnamefont {Amsha}\ \bibnamefont {Proag}}, \bibinfo {author} {\bibfnamefont {Guillaume}\ \bibnamefont {Allio}}, \bibinfo {author} {\bibfnamefont {Bertrand}\ \bibnamefont {Bénazéraf}}, \bibinfo {author} {\bibfnamefont {Jérôme}\ \bibnamefont {Gros}}, \ and\ \bibinfo {author} {\bibfnamefont {Magali}\ \bibnamefont {Suzanne}},\ }\bibfield  {title} {\enquote {\bibinfo {title} {Force-generating apoptotic cells orchestrate avian neural tube bending},}\ }\href {\doibase 10.1016/j.devcel.2022.02.020} {\bibfield  {journal} {\bibinfo  {journal} {Developmental Cell}\ }\textbf {\bibinfo {volume} {57}},\ \bibinfo {pages} {707--718.e6} (\bibinfo {year} {2022})}\BibitemShut {NoStop}%
\bibitem [{\citenamefont {Firmino}\ \emph {et~al.}(2016)\citenamefont {Firmino}, \citenamefont {Rocancourt}, \citenamefont {Saadaoui}, \citenamefont {Moreau},\ and\ \citenamefont {Gros}}]{Firmino2016}%
  \BibitemOpen
  \bibfield  {author} {\bibinfo {author} {\bibfnamefont {Joao}\ \bibnamefont {Firmino}}, \bibinfo {author} {\bibfnamefont {Didier}\ \bibnamefont {Rocancourt}}, \bibinfo {author} {\bibfnamefont {Mehdi}\ \bibnamefont {Saadaoui}}, \bibinfo {author} {\bibfnamefont {Chloe}\ \bibnamefont {Moreau}}, \ and\ \bibinfo {author} {\bibfnamefont {Jerome}\ \bibnamefont {Gros}},\ }\bibfield  {title} {\enquote {\bibinfo {title} {Cell division drives epithelial cell rearrangements during gastrulation in chick},}\ }\href {\doibase 10.1016/j.devcel.2016.01.007} {\bibfield  {journal} {\bibinfo  {journal} {Developmental Cell}\ }\textbf {\bibinfo {volume} {36}},\ \bibinfo {pages} {249--261} (\bibinfo {year} {2016})}\BibitemShut {NoStop}%
\bibitem [{\citenamefont {Matoz-Fernandez}\ \emph {et~al.}(2017)\citenamefont {Matoz-Fernandez}, \citenamefont {Agoritsas}, \citenamefont {Barrat}, \citenamefont {Bertin},\ and\ \citenamefont {Martens}}]{Matoz-Fernandez2017}%
  \BibitemOpen
  \bibfield  {author} {\bibinfo {author} {\bibfnamefont {D.A.}\ \bibnamefont {Matoz-Fernandez}}, \bibinfo {author} {\bibfnamefont {Elisabeth}\ \bibnamefont {Agoritsas}}, \bibinfo {author} {\bibfnamefont {Jean-Louis}\ \bibnamefont {Barrat}}, \bibinfo {author} {\bibfnamefont {Eric}\ \bibnamefont {Bertin}}, \ and\ \bibinfo {author} {\bibfnamefont {Kirsten}\ \bibnamefont {Martens}},\ }\bibfield  {title} {\enquote {\bibinfo {title} {Nonlinear rheology in a model biological tissue},}\ }\href {\doibase 10.1103/PhysRevLett.118.158105} {\bibfield  {journal} {\bibinfo  {journal} {Physical Review Letters}\ }\textbf {\bibinfo {volume} {118}},\ \bibinfo {pages} {158105} (\bibinfo {year} {2017})}\BibitemShut {NoStop}%
\bibitem [{\citenamefont {Bocanegra-Moreno}\ \emph {et~al.}(2023)\citenamefont {Bocanegra-Moreno}, \citenamefont {Singh}, \citenamefont {Hannezo}, \citenamefont {Zagorski},\ and\ \citenamefont {Kicheva}}]{BocanegraMoreno2023}%
  \BibitemOpen
  \bibfield  {author} {\bibinfo {author} {\bibfnamefont {Laura}\ \bibnamefont {Bocanegra-Moreno}}, \bibinfo {author} {\bibfnamefont {Amrita}\ \bibnamefont {Singh}}, \bibinfo {author} {\bibfnamefont {Edouard}\ \bibnamefont {Hannezo}}, \bibinfo {author} {\bibfnamefont {Marcin}\ \bibnamefont {Zagorski}}, \ and\ \bibinfo {author} {\bibfnamefont {Anna}\ \bibnamefont {Kicheva}},\ }\bibfield  {title} {\enquote {\bibinfo {title} {Cell cycle dynamics control fluidity of the developing mouse neuroepithelium},}\ }\href {\doibase 10.1038/s41567-023-01977-w} {\bibfield  {journal} {\bibinfo  {journal} {Nature Physics}\ }\textbf {\bibinfo {volume} {19}},\ \bibinfo {pages} {1050--1058} (\bibinfo {year} {2023})}\BibitemShut {NoStop}%
\bibitem [{\citenamefont {Angelini}\ \emph {et~al.}(2011)\citenamefont {Angelini}, \citenamefont {Hannezo}, \citenamefont {Trepat}, \citenamefont {Marquez}, \citenamefont {Fredberg},\ and\ \citenamefont {Weitz}}]{Angelini2011}%
  \BibitemOpen
  \bibfield  {author} {\bibinfo {author} {\bibfnamefont {T.~E.}\ \bibnamefont {Angelini}}, \bibinfo {author} {\bibfnamefont {E.}~\bibnamefont {Hannezo}}, \bibinfo {author} {\bibfnamefont {X.}~\bibnamefont {Trepat}}, \bibinfo {author} {\bibfnamefont {M.}~\bibnamefont {Marquez}}, \bibinfo {author} {\bibfnamefont {J.~J.}\ \bibnamefont {Fredberg}}, \ and\ \bibinfo {author} {\bibfnamefont {D.~A.}\ \bibnamefont {Weitz}},\ }\bibfield  {title} {\enquote {\bibinfo {title} {Glass-like dynamics of collective cell migration},}\ }\href {\doibase 10.1073/pnas.1010059108} {\bibfield  {journal} {\bibinfo  {journal} {Proceedings of the National Academy of Sciences}\ }\textbf {\bibinfo {volume} {108}},\ \bibinfo {pages} {4714–4719} (\bibinfo {year} {2011})}\BibitemShut {NoStop}%
\bibitem [{\citenamefont {Park}\ \emph {et~al.}(2015)\citenamefont {Park}, \citenamefont {Kim}, \citenamefont {Bi}, \citenamefont {Mitchel}, \citenamefont {Qazvini}, \citenamefont {Tantisira}, \citenamefont {Park}, \citenamefont {McGill}, \citenamefont {Kim}, \citenamefont {Gweon},\ and\ \citenamefont {et~al.}}]{Park2015}%
  \BibitemOpen
  \bibfield  {author} {\bibinfo {author} {\bibfnamefont {Jin-Ah}\ \bibnamefont {Park}}, \bibinfo {author} {\bibfnamefont {Jae~Hun}\ \bibnamefont {Kim}}, \bibinfo {author} {\bibfnamefont {Dapeng}\ \bibnamefont {Bi}}, \bibinfo {author} {\bibfnamefont {Jennifer~A.}\ \bibnamefont {Mitchel}}, \bibinfo {author} {\bibfnamefont {Nader~Taheri}\ \bibnamefont {Qazvini}}, \bibinfo {author} {\bibfnamefont {Kelan}\ \bibnamefont {Tantisira}}, \bibinfo {author} {\bibfnamefont {Chan~Young}\ \bibnamefont {Park}}, \bibinfo {author} {\bibfnamefont {Maureen}\ \bibnamefont {McGill}}, \bibinfo {author} {\bibfnamefont {Sae-Hoon}\ \bibnamefont {Kim}}, \bibinfo {author} {\bibfnamefont {Bomi}\ \bibnamefont {Gweon}}, \ and\ \bibinfo {author} {\bibnamefont {et~al.}},\ }\bibfield  {title} {\enquote {\bibinfo {title} {Unjamming and cell shape in the asthmatic airway epithelium},}\ }\href {\doibase 10.1038/nmat4357} {\bibfield  {journal} {\bibinfo  {journal} {Nature Materials}\ }\textbf {\bibinfo {volume} {14}},\ \bibinfo {pages} {1040–1048} (\bibinfo {year} {2015})}\BibitemShut {NoStop}%
\bibitem [{\citenamefont {Schotz}\ \emph {et~al.}(2013)\citenamefont {Schotz}, \citenamefont {Lanio}, \citenamefont {Talbot},\ and\ \citenamefont {Manning}}]{Schotz2013}%
  \BibitemOpen
  \bibfield  {author} {\bibinfo {author} {\bibfnamefont {E.-M.}\ \bibnamefont {Schotz}}, \bibinfo {author} {\bibfnamefont {M.}~\bibnamefont {Lanio}}, \bibinfo {author} {\bibfnamefont {J.~A.}\ \bibnamefont {Talbot}}, \ and\ \bibinfo {author} {\bibfnamefont {M.~L.}\ \bibnamefont {Manning}},\ }\bibfield  {title} {\enquote {\bibinfo {title} {Glassy dynamics in three-dimensional embryonic tissues},}\ }\href {\doibase 10.1098/rsif.2013.0726} {\bibfield  {journal} {\bibinfo  {journal} {Journal of The Royal Society Interface}\ }\textbf {\bibinfo {volume} {10}},\ \bibinfo {pages} {20130726–20130726} (\bibinfo {year} {2013})}\BibitemShut {NoStop}%
\bibitem [{\citenamefont {Farhadifar}\ \emph {et~al.}(2007)\citenamefont {Farhadifar}, \citenamefont {R{\" o}per}, \citenamefont {Aigouy}, \citenamefont {Eaton},\ and\ \citenamefont {J{\" u}licher}}]{Farhadifar2007}%
  \BibitemOpen
  \bibfield  {author} {\bibinfo {author} {\bibfnamefont {Reza}\ \bibnamefont {Farhadifar}}, \bibinfo {author} {\bibfnamefont {Jens-Christian}\ \bibnamefont {R{\" o}per}}, \bibinfo {author} {\bibfnamefont {Benoit}\ \bibnamefont {Aigouy}}, \bibinfo {author} {\bibfnamefont {Suzanne}\ \bibnamefont {Eaton}}, \ and\ \bibinfo {author} {\bibfnamefont {Frank}\ \bibnamefont {J{\" u}licher}},\ }\bibfield  {title} {\enquote {\bibinfo {title} {The influence of cell mechanics, cell-cell interactions, and proliferation on epithelial packing},}\ }\href {\doibase 10.1016/j.cub.2007.11.049} {\bibfield  {journal} {\bibinfo  {journal} {Current Biology}\ }\textbf {\bibinfo {volume} {17}},\ \bibinfo {pages} {2095–2104} (\bibinfo {year} {2007})}\BibitemShut {NoStop}%
\bibitem [{\citenamefont {Mayer}\ \emph {et~al.}(2010)\citenamefont {Mayer}, \citenamefont {Depken}, \citenamefont {Bois}, \citenamefont {Jülicher},\ and\ \citenamefont {Grill}}]{mayer_anisotropies_2010}%
  \BibitemOpen
  \bibfield  {author} {\bibinfo {author} {\bibfnamefont {Mirjam}\ \bibnamefont {Mayer}}, \bibinfo {author} {\bibfnamefont {Martin}\ \bibnamefont {Depken}}, \bibinfo {author} {\bibfnamefont {Justin~S.}\ \bibnamefont {Bois}}, \bibinfo {author} {\bibfnamefont {Frank}\ \bibnamefont {Jülicher}}, \ and\ \bibinfo {author} {\bibfnamefont {Stephan~W.}\ \bibnamefont {Grill}},\ }\bibfield  {title} {\enquote {\bibinfo {title} {Anisotropies in cortical tension reveal the physical basis of polarizing cortical flows},}\ }\href {\doibase 10.1038/nature09376} {\bibfield  {journal} {\bibinfo  {journal} {Nature}\ }\textbf {\bibinfo {volume} {467}},\ \bibinfo {pages} {617--621} (\bibinfo {year} {2010})}\BibitemShut {NoStop}%
\bibitem [{\citenamefont {Bonnet}\ \emph {et~al.}(2012)\citenamefont {Bonnet}, \citenamefont {Marcq}, \citenamefont {Bosveld}, \citenamefont {Fetler}, \citenamefont {Bellaïche},\ and\ \citenamefont {Graner}}]{Bonnet2012}%
  \BibitemOpen
  \bibfield  {author} {\bibinfo {author} {\bibfnamefont {Isabelle}\ \bibnamefont {Bonnet}}, \bibinfo {author} {\bibfnamefont {Philippe}\ \bibnamefont {Marcq}}, \bibinfo {author} {\bibfnamefont {Floris}\ \bibnamefont {Bosveld}}, \bibinfo {author} {\bibfnamefont {Luc}\ \bibnamefont {Fetler}}, \bibinfo {author} {\bibfnamefont {Yohanns}\ \bibnamefont {Bellaïche}}, \ and\ \bibinfo {author} {\bibfnamefont {François}\ \bibnamefont {Graner}},\ }\bibfield  {title} {\enquote {\bibinfo {title} {Mechanical state, material properties and continuous description of an epithelial tissue},}\ }\href {\doibase 10.1098/rsif.2012.0263} {\bibfield  {journal} {\bibinfo  {journal} {Journal of The Royal Society Interface}\ }\textbf {\bibinfo {volume} {9}},\ \bibinfo {pages} {2614--2623} (\bibinfo {year} {2012})}\BibitemShut {NoStop}%
\bibitem [{\citenamefont {Kasza}\ \emph {et~al.}(2014)\citenamefont {Kasza}, \citenamefont {Farrell},\ and\ \citenamefont {Zallen}}]{Kasza2014}%
  \BibitemOpen
  \bibfield  {author} {\bibinfo {author} {\bibfnamefont {Karen~E.}\ \bibnamefont {Kasza}}, \bibinfo {author} {\bibfnamefont {Dene~L.}\ \bibnamefont {Farrell}}, \ and\ \bibinfo {author} {\bibfnamefont {Jennifer~A.}\ \bibnamefont {Zallen}},\ }\bibfield  {title} {\enquote {\bibinfo {title} {Spatiotemporal control of epithelial remodeling by regulated myosin phosphorylation},}\ }\href {\doibase 10.1073/pnas.1400520111} {\bibfield  {journal} {\bibinfo  {journal} {Proceedings of the National Academy of Sciences}\ }\textbf {\bibinfo {volume} {111}},\ \bibinfo {pages} {11732–11737} (\bibinfo {year} {2014})}\BibitemShut {NoStop}%
\bibitem [{\citenamefont {Piscitello-Gómez}\ \emph {et~al.}(2023)\citenamefont {Piscitello-Gómez}, \citenamefont {Gruber}, \citenamefont {Krishna}, \citenamefont {Duclut}, \citenamefont {Modes}, \citenamefont {Popović}, \citenamefont {Jülicher}, \citenamefont {Dye},\ and\ \citenamefont {Eaton}}]{Piscitello2022}%
  \BibitemOpen
  \bibfield  {author} {\bibinfo {author} {\bibfnamefont {Romina}\ \bibnamefont {Piscitello-Gómez}}, \bibinfo {author} {\bibfnamefont {Franz~S}\ \bibnamefont {Gruber}}, \bibinfo {author} {\bibfnamefont {Abhijeet}\ \bibnamefont {Krishna}}, \bibinfo {author} {\bibfnamefont {Charlie}\ \bibnamefont {Duclut}}, \bibinfo {author} {\bibfnamefont {Carl~D}\ \bibnamefont {Modes}}, \bibinfo {author} {\bibfnamefont {Marko}\ \bibnamefont {Popović}}, \bibinfo {author} {\bibfnamefont {Frank}\ \bibnamefont {Jülicher}}, \bibinfo {author} {\bibfnamefont {Natalie~A}\ \bibnamefont {Dye}}, \ and\ \bibinfo {author} {\bibfnamefont {Suzanne}\ \bibnamefont {Eaton}},\ }\bibfield  {title} {\enquote {\bibinfo {title} {Core pcp mutations affect short-time mechanical properties but not tissue morphogenesis in the \textit{Drosophila} pupal wing},}\ }\href {\doibase 10.7554/eLife.85581} {\bibfield  {journal} {\bibinfo  {journal} {eLife}\ }\textbf {\bibinfo {volume} {12}},\ \bibinfo {pages} {e85581} (\bibinfo {year} {2023})}\BibitemShut {NoStop}%
\bibitem [{\citenamefont {Dye}\ \emph {et~al.}(2021)\citenamefont {Dye}, \citenamefont {Popovi{\' c}}, \citenamefont {Iyer}, \citenamefont {Fuhrmann}, \citenamefont {Piscitello-G{\' o}mez}, \citenamefont {Eaton},\ and\ \citenamefont {J{\" u}licher}}]{Dye2021}%
  \BibitemOpen
  \bibfield  {author} {\bibinfo {author} {\bibfnamefont {Natalie~A}\ \bibnamefont {Dye}}, \bibinfo {author} {\bibfnamefont {Marko}\ \bibnamefont {Popovi{\' c}}}, \bibinfo {author} {\bibfnamefont {K~Venkatesan}\ \bibnamefont {Iyer}}, \bibinfo {author} {\bibfnamefont {Jana~F}\ \bibnamefont {Fuhrmann}}, \bibinfo {author} {\bibfnamefont {Romina}\ \bibnamefont {Piscitello-G{\' o}mez}}, \bibinfo {author} {\bibfnamefont {Suzanne}\ \bibnamefont {Eaton}}, \ and\ \bibinfo {author} {\bibfnamefont {Frank}\ \bibnamefont {J{\" u}licher}},\ }\bibfield  {title} {\enquote {\bibinfo {title} {Self-organized patterning of cell morphology via mechanosensitive feedback},}\ }\href {\doibase 10.7554/eLife.57964} {\bibfield  {journal} {\bibinfo  {journal} {eLife}\ }\textbf {\bibinfo {volume} {10}},\ \bibinfo {pages} {e57964} (\bibinfo {year} {2021})}\BibitemShut {NoStop}%
\bibitem [{\citenamefont {Moriel}\ \emph {et~al.}(2020)\citenamefont {Moriel}, \citenamefont {Lubomirsky}, \citenamefont {Lerner},\ and\ \citenamefont {Bouchbinder}}]{Moriel2020}%
  \BibitemOpen
  \bibfield  {author} {\bibinfo {author} {\bibfnamefont {Avraham}\ \bibnamefont {Moriel}}, \bibinfo {author} {\bibfnamefont {Yuri}\ \bibnamefont {Lubomirsky}}, \bibinfo {author} {\bibfnamefont {Edan}\ \bibnamefont {Lerner}}, \ and\ \bibinfo {author} {\bibfnamefont {Eran}\ \bibnamefont {Bouchbinder}},\ }\bibfield  {title} {\enquote {\bibinfo {title} {Extracting the properties of quasilocalized modes in computer glasses: Long-range continuum fields, contour integrals, and boundary effects},}\ }\href {\doibase 10.1103/PhysRevE.102.033008} {\bibfield  {journal} {\bibinfo  {journal} {Physical Review E}\ }\textbf {\bibinfo {volume} {102}},\ \bibinfo {pages} {033008} (\bibinfo {year} {2020})}\BibitemShut {NoStop}%
\bibitem [{\citenamefont {Chacko}\ \emph {et~al.}(2021)\citenamefont {Chacko}, \citenamefont {Landes}, \citenamefont {Biroli}, \citenamefont {Dauchot}, \citenamefont {Liu},\ and\ \citenamefont {Reichman}}]{Chacko2021}%
  \BibitemOpen
  \bibfield  {author} {\bibinfo {author} {\bibfnamefont {Rahul~N.}\ \bibnamefont {Chacko}}, \bibinfo {author} {\bibfnamefont {François~P.}\ \bibnamefont {Landes}}, \bibinfo {author} {\bibfnamefont {Giulio}\ \bibnamefont {Biroli}}, \bibinfo {author} {\bibfnamefont {Olivier}\ \bibnamefont {Dauchot}}, \bibinfo {author} {\bibfnamefont {Andrea~J.}\ \bibnamefont {Liu}}, \ and\ \bibinfo {author} {\bibfnamefont {David~R.}\ \bibnamefont {Reichman}},\ }\bibfield  {title} {\enquote {\bibinfo {title} {Elastoplasticity mediates dynamical heterogeneity below the mode coupling temperature},}\ }\href {\doibase 10.1103/PhysRevLett.127.048002} {\bibfield  {journal} {\bibinfo  {journal} {Physical Review Letters}\ }\textbf {\bibinfo {volume} {127}},\ \bibinfo {pages} {048002} (\bibinfo {year} {2021})}\BibitemShut {NoStop}%
\bibitem [{\citenamefont {Moriel}\ \emph {et~al.}(2024)\citenamefont {Moriel}, \citenamefont {Richard}, \citenamefont {Lerner},\ and\ \citenamefont {Bouchbinder}}]{Moriel2024}%
  \BibitemOpen
  \bibfield  {author} {\bibinfo {author} {\bibfnamefont {Avraham}\ \bibnamefont {Moriel}}, \bibinfo {author} {\bibfnamefont {David}\ \bibnamefont {Richard}}, \bibinfo {author} {\bibfnamefont {Edan}\ \bibnamefont {Lerner}}, \ and\ \bibinfo {author} {\bibfnamefont {Eran}\ \bibnamefont {Bouchbinder}},\ }\bibfield  {title} {\enquote {\bibinfo {title} {Elementary processes in dilatational plasticity of glasses},}\ }\href {\doibase 10.1103/PhysRevResearch.6.023167} {\bibfield  {journal} {\bibinfo  {journal} {Physical Review Research}\ }\textbf {\bibinfo {volume} {6}},\ \bibinfo {pages} {023167} (\bibinfo {year} {2024})}\BibitemShut {NoStop}%
\bibitem [{\citenamefont {Fuhrmann}\ \emph {et~al.}(2023)\citenamefont {Fuhrmann}, \citenamefont {Krishna}, \citenamefont {Paijmans}, \citenamefont {Duclut}, \citenamefont {Eaton}, \citenamefont {Popovi\'c}, \citenamefont {J\"ulicher}, \citenamefont {Modes},\ and\ \citenamefont {Dye}}]{Fuhrmann2023}%
  \BibitemOpen
  \bibfield  {author} {\bibinfo {author} {\bibfnamefont {Jana~F.}\ \bibnamefont {Fuhrmann}}, \bibinfo {author} {\bibfnamefont {Abhijeet}\ \bibnamefont {Krishna}}, \bibinfo {author} {\bibfnamefont {Joris}\ \bibnamefont {Paijmans}}, \bibinfo {author} {\bibfnamefont {Charlie}\ \bibnamefont {Duclut}}, \bibinfo {author} {\bibfnamefont {Suzanne}\ \bibnamefont {Eaton}}, \bibinfo {author} {\bibfnamefont {Marko}\ \bibnamefont {Popovi\'c}}, \bibinfo {author} {\bibfnamefont {Frank}\ \bibnamefont {J\"ulicher}}, \bibinfo {author} {\bibfnamefont {Carl~D.}\ \bibnamefont {Modes}}, \ and\ \bibinfo {author} {\bibfnamefont {Natalie~A.}\ \bibnamefont {Dye}},\ }\bibfield  {title} {\enquote {\bibinfo {title} {Active shape programming drives drosophila wing disc eversion},}\ }\href {\doibase 10.1101/2023.12.23.573034} {\  (\bibinfo {year} {2023}),\ 10.1101/2023.12.23.573034}\BibitemShut {NoStop}%
\bibitem [{\citenamefont {Merkel}\ \emph {et~al.}(2017)\citenamefont {Merkel}, \citenamefont {Etournay}, \citenamefont {Popovi{\' c}}, \citenamefont {Salbreux}, \citenamefont {Eaton},\ and\ \citenamefont {J{\" u}licher}}]{Merkel2017}%
  \BibitemOpen
  \bibfield  {author} {\bibinfo {author} {\bibfnamefont {Matthias}\ \bibnamefont {Merkel}}, \bibinfo {author} {\bibfnamefont {Rapha{\" e}l}\ \bibnamefont {Etournay}}, \bibinfo {author} {\bibfnamefont {Marko}\ \bibnamefont {Popovi{\' c}}}, \bibinfo {author} {\bibfnamefont {Guillaume}\ \bibnamefont {Salbreux}}, \bibinfo {author} {\bibfnamefont {Suzanne}\ \bibnamefont {Eaton}}, \ and\ \bibinfo {author} {\bibfnamefont {Frank}\ \bibnamefont {J{\" u}licher}},\ }\bibfield  {title} {\enquote {\bibinfo {title} {Triangles bridge the scales: Quantifying cellular contributions to tissue deformation},}\ }\href {\doibase 10.1103/PhysRevE.95.032401} {\bibfield  {journal} {\bibinfo  {journal} {Physical Review E}\ }\textbf {\bibinfo {volume} {95}} (\bibinfo {year} {2017}),\ 10.1103/PhysRevE.95.032401}\BibitemShut {NoStop}%
\bibitem [{\citenamefont {Eshelby}(1957)}]{Eshelby1957}%
  \BibitemOpen
  \bibfield  {author} {\bibinfo {author} {\bibfnamefont {John~Douglas}\ \bibnamefont {Eshelby}},\ }\bibfield  {title} {\enquote {\bibinfo {title} {The determination of the elastic field of an ellipsoidal inclusion, and related problems},}\ }\href@noop {} {\bibfield  {journal} {\bibinfo  {journal} {Proceedings of the royal society of London. Series A. Mathematical and physical sciences}\ }\textbf {\bibinfo {volume} {241}},\ \bibinfo {pages} {376--396} (\bibinfo {year} {1957})}\BibitemShut {NoStop}%
\bibitem [{\citenamefont {Nicolas}\ \emph {et~al.}(2018)\citenamefont {Nicolas}, \citenamefont {Ferrero}, \citenamefont {Martens},\ and\ \citenamefont {Barrat}}]{Nicolas2018}%
  \BibitemOpen
  \bibfield  {author} {\bibinfo {author} {\bibfnamefont {Alexandre}\ \bibnamefont {Nicolas}}, \bibinfo {author} {\bibfnamefont {Ezequiel~E.}\ \bibnamefont {Ferrero}}, \bibinfo {author} {\bibfnamefont {Kirsten}\ \bibnamefont {Martens}}, \ and\ \bibinfo {author} {\bibfnamefont {Jean-Louis}\ \bibnamefont {Barrat}},\ }\bibfield  {title} {\enquote {\bibinfo {title} {Deformation and flow of amorphous solids: Insights from elastoplastic models},}\ }\href {\doibase 10.1103/RevModPhys.90.045006} {\bibfield  {journal} {\bibinfo  {journal} {Reviews of Modern Physics}\ }\textbf {\bibinfo {volume} {90}} (\bibinfo {year} {2018}),\ 10.1103/RevModPhys.90.045006}\BibitemShut {NoStop}%
\bibitem [{Note1()}]{Note1}%
  \BibitemOpen
  \bibinfo {note} {We consider only force dipoles that do not exert a net torque.}\BibitemShut {Stop}%
\bibitem [{\citenamefont {Etournay}\ \emph {et~al.}(2016)\citenamefont {Etournay}, \citenamefont {Merkel}, \citenamefont {Popovi{\' c}}, \citenamefont {Brandl}, \citenamefont {Dye}, \citenamefont {Aigouy}, \citenamefont {Salbreux}, \citenamefont {Eaton},\ and\ \citenamefont {J{\" u}licher}}]{Etournay2016}%
  \BibitemOpen
  \bibfield  {author} {\bibinfo {author} {\bibfnamefont {Rapha{\" e}l}\ \bibnamefont {Etournay}}, \bibinfo {author} {\bibfnamefont {Matthias}\ \bibnamefont {Merkel}}, \bibinfo {author} {\bibfnamefont {Marko}\ \bibnamefont {Popovi{\' c}}}, \bibinfo {author} {\bibfnamefont {Holger}\ \bibnamefont {Brandl}}, \bibinfo {author} {\bibfnamefont {Natalie~A}\ \bibnamefont {Dye}}, \bibinfo {author} {\bibfnamefont {Benoit}\ \bibnamefont {Aigouy}}, \bibinfo {author} {\bibfnamefont {Guillaume}\ \bibnamefont {Salbreux}}, \bibinfo {author} {\bibfnamefont {Suzanne}\ \bibnamefont {Eaton}}, \ and\ \bibinfo {author} {\bibfnamefont {Frank}\ \bibnamefont {J{\" u}licher}},\ }\bibfield  {title} {\enquote {\bibinfo {title} {Tissueminer: A multiscale analysis toolkit to quantify how cellular processes create tissue dynamics},}\ }\href {\doibase 10.7554/eLife.14334} {\bibfield  {journal} {\bibinfo  {journal} {eLife}\ }\textbf {\bibinfo {volume} {5}} (\bibinfo {year} {2016}),\ 10.7554/eLife.14334}\BibitemShut {NoStop}%
\bibitem [{\citenamefont {Strzyz}\ \emph {et~al.}(2016)\citenamefont {Strzyz}, \citenamefont {Matejcic},\ and\ \citenamefont {Norden}}]{Jeon2016}%
  \BibitemOpen
  \bibfield  {author} {\bibinfo {author} {\bibfnamefont {P.~J.}\ \bibnamefont {Strzyz}}, \bibinfo {author} {\bibfnamefont {M.}~\bibnamefont {Matejcic}}, \ and\ \bibinfo {author} {\bibfnamefont {C.}~\bibnamefont {Norden}},\ }\bibfield  {title} {\enquote {\bibinfo {title} {Chapter {Three} - {Heterogeneity}, {Cell} {Biology} and {Tissue} {Mechanics} of {Pseudostratified} {Epithelia}: {Coordination} of {Cell} {Divisions} and {Growth} in {Tightly} {Packed} {Tissues}},}\ }in\ \href {\doibase https://doi.org/10.1016/bs.ircmb.2016.02.004} {\emph {\bibinfo {booktitle} {International {Review} of {Cell} and {Molecular} {Biology}}}},\ \bibinfo {series} {International {Review} of {Cell} and {Molecular} {Biology}}, Vol.\ \bibinfo {volume} {325},\ \bibinfo {editor} {edited by\ \bibinfo {editor} {\bibfnamefont {Kwang~W.}\ \bibnamefont {Jeon}}}\ (\bibinfo  {publisher} {Academic Press},\ \bibinfo {year} {2016})\ pp.\ \bibinfo {pages} {89--118},\ \bibinfo {note} {iSSN: 1937-6448}\BibitemShut {NoStop}%
\bibitem [{\citenamefont {Hertwig}(1884)}]{Hertwig1884}%
  \BibitemOpen
  \bibfield  {author} {\bibinfo {author} {\bibfnamefont {O.}~\bibnamefont {Hertwig}},\ }\bibfield  {title} {\enquote {\bibinfo {title} {Das problem der befruchtung und der isotropie des eies: eine theorie der vererbung},}\ }\href@noop {} {\ \bibinfo {series} {Jenaische Zeitschrift f\"ur Naturwissenschaften} (\bibinfo {year} {1884})}\BibitemShut {NoStop}%
\bibitem [{\citenamefont {Th\'ery}\ \emph {et~al.}(2007)\citenamefont {Th\'ery}, \citenamefont {Jim\'enez-Dalmaroni}, \citenamefont {Racine}, \citenamefont {Bornens},\ and\ \citenamefont {J{\" u}licher}}]{Thery2007}%
  \BibitemOpen
  \bibfield  {author} {\bibinfo {author} {\bibfnamefont {Manuel}\ \bibnamefont {Th\'ery}}, \bibinfo {author} {\bibfnamefont {Andrea}\ \bibnamefont {Jim\'enez-Dalmaroni}}, \bibinfo {author} {\bibfnamefont {Victor}\ \bibnamefont {Racine}}, \bibinfo {author} {\bibfnamefont {Michel}\ \bibnamefont {Bornens}}, \ and\ \bibinfo {author} {\bibfnamefont {Frank}\ \bibnamefont {J{\" u}licher}},\ }\bibfield  {title} {\enquote {\bibinfo {title} {Experimental and theoretical study of mitotic spindle orientation},}\ }\href {\doibase 10.1038/nature05786} {\bibfield  {journal} {\bibinfo  {journal} {Nature}\ }\textbf {\bibinfo {volume} {447}},\ \bibinfo {pages} {493--496} (\bibinfo {year} {2007})}\BibitemShut {NoStop}%
\bibitem [{\citenamefont {Lisica}\ \emph {et~al.}(2022)\citenamefont {Lisica}, \citenamefont {Fouchard}, \citenamefont {Kelkar}, \citenamefont {Wyatt}, \citenamefont {Duque}, \citenamefont {Ndiaye}, \citenamefont {Bonfanti}, \citenamefont {Baum}, \citenamefont {Kabla},\ and\ \citenamefont {Charras}}]{Lisica2022}%
  \BibitemOpen
  \bibfield  {author} {\bibinfo {author} {\bibfnamefont {Ana}\ \bibnamefont {Lisica}}, \bibinfo {author} {\bibfnamefont {Jonathan}\ \bibnamefont {Fouchard}}, \bibinfo {author} {\bibfnamefont {Manasi}\ \bibnamefont {Kelkar}}, \bibinfo {author} {\bibfnamefont {Tom P.~J.}\ \bibnamefont {Wyatt}}, \bibinfo {author} {\bibfnamefont {Julia}\ \bibnamefont {Duque}}, \bibinfo {author} {\bibfnamefont {Anne-Betty}\ \bibnamefont {Ndiaye}}, \bibinfo {author} {\bibfnamefont {Alessandra}\ \bibnamefont {Bonfanti}}, \bibinfo {author} {\bibfnamefont {Buzz}\ \bibnamefont {Baum}}, \bibinfo {author} {\bibfnamefont {Alexandre~J.}\ \bibnamefont {Kabla}}, \ and\ \bibinfo {author} {\bibfnamefont {Guillaume~T.}\ \bibnamefont {Charras}},\ }\bibfield  {title} {\enquote {\bibinfo {title} {Tension at intercellular junctions is necessary for accurate orientation of cell division in the epithelium plane},}\ }\href {\doibase 10.1073/pnas.2201600119} {\bibfield  {journal} {\bibinfo  {journal} {Proceedings of the National Academy of Sciences}\ }\textbf {\bibinfo {volume} {119}},\ \bibinfo {pages} {e2201600119} (\bibinfo {year} {2022})}\BibitemShut {NoStop}%
\bibitem [{\citenamefont {Middelkoop}\ \emph {et~al.}(2023)\citenamefont {Middelkoop}, \citenamefont {Neipel}, \citenamefont {Cornell}, \citenamefont {Naumann}, \citenamefont {Pimpale}, \citenamefont {Jülicher},\ and\ \citenamefont {Grill}}]{Middelkoop2023}%
  \BibitemOpen
  \bibfield  {author} {\bibinfo {author} {\bibfnamefont {Teije~C.}\ \bibnamefont {Middelkoop}}, \bibinfo {author} {\bibfnamefont {Jonas}\ \bibnamefont {Neipel}}, \bibinfo {author} {\bibfnamefont {Caitlin~E.}\ \bibnamefont {Cornell}}, \bibinfo {author} {\bibfnamefont {Ronald}\ \bibnamefont {Naumann}}, \bibinfo {author} {\bibfnamefont {Lokesh~G.}\ \bibnamefont {Pimpale}}, \bibinfo {author} {\bibfnamefont {Frank}\ \bibnamefont {Jülicher}}, \ and\ \bibinfo {author} {\bibfnamefont {Stephan~W.}\ \bibnamefont {Grill}},\ }\bibfield  {title} {\enquote {\bibinfo {title} {A cytokinetic ring-driven cell rotation achieves hertwig’s rule in early development},}\ }\href {\doibase 10.1101/2023.06.23.546115} {\  (\bibinfo {year} {2023}),\ 10.1101/2023.06.23.546115}\BibitemShut {NoStop}%
\bibitem [{\citenamefont {Sun}\ \emph {et~al.}(2021)\citenamefont {Sun}, \citenamefont {Song}, \citenamefont {Teng}, \citenamefont {Chen}, \citenamefont {Dai}, \citenamefont {Ma}, \citenamefont {Zhang},\ and\ \citenamefont {Pastor-Pareja}}]{Sun2021}%
  \BibitemOpen
  \bibfield  {author} {\bibinfo {author} {\bibfnamefont {Tianhui}\ \bibnamefont {Sun}}, \bibinfo {author} {\bibfnamefont {Yuzhao}\ \bibnamefont {Song}}, \bibinfo {author} {\bibfnamefont {Dequn}\ \bibnamefont {Teng}}, \bibinfo {author} {\bibfnamefont {Yanan}\ \bibnamefont {Chen}}, \bibinfo {author} {\bibfnamefont {Jianli}\ \bibnamefont {Dai}}, \bibinfo {author} {\bibfnamefont {Mengqi}\ \bibnamefont {Ma}}, \bibinfo {author} {\bibfnamefont {Wei}\ \bibnamefont {Zhang}}, \ and\ \bibinfo {author} {\bibfnamefont {Jos\' e~C.}\ \bibnamefont {Pastor-Pareja}},\ }\bibfield  {title} {\enquote {\bibinfo {title} {Atypical laminin spots and pull-generated microtubule-actin projections mediate drosophila wing adhesion},}\ }\href {\doibase 10.1016/j.celrep.2021.109667} {\bibfield  {journal} {\bibinfo  {journal} {Cell Reports}\ }\textbf {\bibinfo {volume} {36}},\ \bibinfo {pages} {109667} (\bibinfo {year} {2021})}\BibitemShut {NoStop}%
\bibitem [{\citenamefont {Dyre}(2023)}]{Dyre2023}%
  \BibitemOpen
  \bibfield  {author} {\bibinfo {author} {\bibfnamefont {Jeppe~C.}\ \bibnamefont {Dyre}},\ }\bibfield  {title} {\enquote {\bibinfo {title} {Solid-that-flows picture of glass-forming liquids},}\ }\href {\doibase 10.48550/arXiv.2311.14460} {\bibfield  {journal} {\bibinfo  {journal} {arXiv}\ } (\bibinfo {year} {2023}),\ 10.48550/arXiv.2311.14460}\BibitemShut {NoStop}%
\bibitem [{\citenamefont {Ozawa}\ and\ \citenamefont {Biroli}(2023)}]{Ozawa2023}%
  \BibitemOpen
  \bibfield  {author} {\bibinfo {author} {\bibfnamefont {Misaki}\ \bibnamefont {Ozawa}}\ and\ \bibinfo {author} {\bibfnamefont {Giulio}\ \bibnamefont {Biroli}},\ }\bibfield  {title} {\enquote {\bibinfo {title} {Elasticity, facilitation, and dynamic heterogeneity in glass-forming liquids},}\ }\href {\doibase 10.1103/PhysRevLett.130.138201} {\bibfield  {journal} {\bibinfo  {journal} {Phys. Rev. Lett.}\ }\textbf {\bibinfo {volume} {130}},\ \bibinfo {pages} {138201} (\bibinfo {year} {2023})}\BibitemShut {NoStop}%
\bibitem [{\citenamefont {Tahaei}\ \emph {et~al.}(2023)\citenamefont {Tahaei}, \citenamefont {Biroli}, \citenamefont {Ozawa}, \citenamefont {Popovi\ifmmode~\acute{c}\else \'{c}\fi{}},\ and\ \citenamefont {Wyart}}]{Tahaei2023}%
  \BibitemOpen
  \bibfield  {author} {\bibinfo {author} {\bibfnamefont {Ali}\ \bibnamefont {Tahaei}}, \bibinfo {author} {\bibfnamefont {Giulio}\ \bibnamefont {Biroli}}, \bibinfo {author} {\bibfnamefont {Misaki}\ \bibnamefont {Ozawa}}, \bibinfo {author} {\bibfnamefont {Marko}\ \bibnamefont {Popovi\ifmmode~\acute{c}\else \'{c}\fi{}}}, \ and\ \bibinfo {author} {\bibfnamefont {Matthieu}\ \bibnamefont {Wyart}},\ }\bibfield  {title} {\enquote {\bibinfo {title} {Scaling description of dynamical heterogeneity and avalanches of relaxation in glass-forming liquids},}\ }\href {\doibase 10.1103/PhysRevX.13.031034} {\bibfield  {journal} {\bibinfo  {journal} {Phys. Rev. X}\ }\textbf {\bibinfo {volume} {13}},\ \bibinfo {pages} {031034} (\bibinfo {year} {2023})}\BibitemShut {NoStop}%
\bibitem [{\citenamefont {Aigouy}\ \emph {et~al.}(2016)\citenamefont {Aigouy}, \citenamefont {Umetsu},\ and\ \citenamefont {Eaton}}]{Aigouy2016}%
  \BibitemOpen
  \bibfield  {author} {\bibinfo {author} {\bibfnamefont {Benoit}\ \bibnamefont {Aigouy}}, \bibinfo {author} {\bibfnamefont {Daiki}\ \bibnamefont {Umetsu}}, \ and\ \bibinfo {author} {\bibfnamefont {Suzanne}\ \bibnamefont {Eaton}},\ }\enquote {\bibinfo {title} {Segmentation and quantitative analysis of epithelial tissues},}\ in\ \href {\doibase 10.1007/978-1-4939-6371-3_13} {\emph {\bibinfo {booktitle} {Drosophila: Methods and Protocols}}},\ \bibinfo {editor} {edited by\ \bibinfo {editor} {\bibfnamefont {Christian}\ \bibnamefont {Dahmann}}}\ (\bibinfo  {publisher} {Springer New York},\ \bibinfo {address} {New York, NY},\ \bibinfo {year} {2016})\ pp.\ \bibinfo {pages} {227--239}\BibitemShut {NoStop}%
\bibitem [{\citenamefont {Dye}\ \emph {et~al.}(2017)\citenamefont {Dye}, \citenamefont {Popovi{\' c}}, \citenamefont {Spannl}, \citenamefont {Etournay}, \citenamefont {Kainm{\" u}ller}, \citenamefont {Ghosh}, \citenamefont {Myers}, \citenamefont {J{\" u}licher},\ and\ \citenamefont {Eaton}}]{Dye2017}%
  \BibitemOpen
  \bibfield  {author} {\bibinfo {author} {\bibfnamefont {Natalie~A.}\ \bibnamefont {Dye}}, \bibinfo {author} {\bibfnamefont {Marko}\ \bibnamefont {Popovi{\' c}}}, \bibinfo {author} {\bibfnamefont {Stephanie}\ \bibnamefont {Spannl}}, \bibinfo {author} {\bibfnamefont {Rapha{\" e}l}\ \bibnamefont {Etournay}}, \bibinfo {author} {\bibfnamefont {Dagmar}\ \bibnamefont {Kainm{\" u}ller}}, \bibinfo {author} {\bibfnamefont {Suhrid}\ \bibnamefont {Ghosh}}, \bibinfo {author} {\bibfnamefont {Eugene~W.}\ \bibnamefont {Myers}}, \bibinfo {author} {\bibfnamefont {Frank}\ \bibnamefont {J{\" u}licher}}, \ and\ \bibinfo {author} {\bibfnamefont {Suzanne}\ \bibnamefont {Eaton}},\ }\bibfield  {title} {\enquote {\bibinfo {title} {Cell dynamics underlying oriented growth of the drosophila wing imaginal disc},}\ }\href {\doibase 10.1242/dev.155069} {\bibfield  {journal} {\bibinfo  {journal} {Development}\ }\textbf {\bibinfo {volume} {144}},\ \bibinfo {pages} {4406–4421} (\bibinfo {year} {2017})}\BibitemShut {NoStop}%
\bibitem [{\citenamefont {Duclut}\ \emph {et~al.}(2021)\citenamefont {Duclut}, \citenamefont {Paijmans}, \citenamefont {Inamdar}, \citenamefont {Modes},\ and\ \citenamefont {Jülicher}}]{Duclut2021}%
  \BibitemOpen
  \bibfield  {author} {\bibinfo {author} {\bibfnamefont {Charlie}\ \bibnamefont {Duclut}}, \bibinfo {author} {\bibfnamefont {Joris}\ \bibnamefont {Paijmans}}, \bibinfo {author} {\bibfnamefont {Mandar~M.}\ \bibnamefont {Inamdar}}, \bibinfo {author} {\bibfnamefont {Carl~D.}\ \bibnamefont {Modes}}, \ and\ \bibinfo {author} {\bibfnamefont {Frank}\ \bibnamefont {Jülicher}},\ }\bibfield  {title} {\enquote {\bibinfo {title} {Nonlinear rheology of cellular networks},}\ }\href {\doibase 10.1016/j.cdev.2021.203746} {\bibfield  {journal} {\bibinfo  {journal} {Cells \& Development}\ }\textbf {\bibinfo {volume} {168}},\ \bibinfo {pages} {203746} (\bibinfo {year} {2021})}\BibitemShut {NoStop}%
\end{thebibliography}%


\begin{thebibliography}{3}%
\makeatletter
\providecommand \@ifxundefined [1]{%
 \@ifx{#1\undefined}
}%
\providecommand \@ifnum [1]{%
 \ifnum #1\expandafter \@firstoftwo
 \else \expandafter \@secondoftwo
 \fi
}%
\providecommand \@ifx [1]{%
 \ifx #1\expandafter \@firstoftwo
 \else \expandafter \@secondoftwo
 \fi
}%
\providecommand \natexlab [1]{#1}%
\providecommand \enquote  [1]{``#1''}%
\providecommand \bibnamefont  [1]{#1}%
\providecommand \bibfnamefont [1]{#1}%
\providecommand \citenamefont [1]{#1}%
\providecommand \href@noop [0]{\@secondoftwo}%
\providecommand \href [0]{\begingroup \@sanitize@url \@href}%
\providecommand \@href[1]{\@@startlink{#1}\@@href}%
\providecommand \@@href[1]{\endgroup#1\@@endlink}%
\providecommand \@sanitize@url [0]{\catcode `\\12\catcode `\$12\catcode `\&12\catcode `\#12\catcode `\^12\catcode `\_12\catcode `\%12\relax}%
\providecommand \@@startlink[1]{}%
\providecommand \@@endlink[0]{}%
\providecommand \url  [0]{\begingroup\@sanitize@url \@url }%
\providecommand \@url [1]{\endgroup\@href {#1}{\urlprefix }}%
\providecommand \urlprefix  [0]{URL }%
\providecommand \Eprint [0]{\href }%
\providecommand \doibase [0]{http://dx.doi.org/}%
\providecommand \selectlanguage [0]{\@gobble}%
\providecommand \bibinfo  [0]{\@secondoftwo}%
\providecommand \bibfield  [0]{\@secondoftwo}%
\providecommand \translation [1]{[#1]}%
\providecommand \BibitemOpen [0]{}%
\providecommand \bibitemStop [0]{}%
\providecommand \bibitemNoStop [0]{.\EOS\space}%
\providecommand \EOS [0]{\spacefactor3000\relax}%
\providecommand \BibitemShut  [1]{\csname bibitem#1\endcsname}%
\let\auto@bib@innerbib\@empty
\bibitem [{\citenamefont {Piscitello-Gómez}\ \emph {et~al.}(2023)\citenamefont {Piscitello-Gómez}, \citenamefont {Gruber}, \citenamefont {Krishna}, \citenamefont {Duclut}, \citenamefont {Modes}, \citenamefont {Popović}, \citenamefont {Jülicher}, \citenamefont {Dye},\ and\ \citenamefont {Eaton}}]{Piscitello2022}%
  \BibitemOpen
  \bibfield  {author} {\bibinfo {author} {\bibfnamefont {Romina}\ \bibnamefont {Piscitello-Gómez}}, \bibinfo {author} {\bibfnamefont {Franz~S}\ \bibnamefont {Gruber}}, \bibinfo {author} {\bibfnamefont {Abhijeet}\ \bibnamefont {Krishna}}, \bibinfo {author} {\bibfnamefont {Charlie}\ \bibnamefont {Duclut}}, \bibinfo {author} {\bibfnamefont {Carl~D}\ \bibnamefont {Modes}}, \bibinfo {author} {\bibfnamefont {Marko}\ \bibnamefont {Popović}}, \bibinfo {author} {\bibfnamefont {Frank}\ \bibnamefont {Jülicher}}, \bibinfo {author} {\bibfnamefont {Natalie~A}\ \bibnamefont {Dye}}, \ and\ \bibinfo {author} {\bibfnamefont {Suzanne}\ \bibnamefont {Eaton}},\ }\bibfield  {title} {\enquote {\bibinfo {title} {Core pcp mutations affect short-time mechanical properties but not tissue morphogenesis in the \textit{Drosophila} pupal wing},}\ }\href {\doibase 10.7554/eLife.85581} {\bibfield  {journal} {\bibinfo  {journal} {eLife}\ }\textbf {\bibinfo {volume} {12}},\ \bibinfo {pages} {e85581} (\bibinfo {year} {2023})}\BibitemShut {NoStop}%
\bibitem [{\citenamefont {Farhadifar}\ \emph {et~al.}(2007)\citenamefont {Farhadifar}, \citenamefont {R{\" o}per}, \citenamefont {Aigouy}, \citenamefont {Eaton},\ and\ \citenamefont {J{\" u}licher}}]{Farhadifar2007}%
  \BibitemOpen
  \bibfield  {author} {\bibinfo {author} {\bibfnamefont {Reza}\ \bibnamefont {Farhadifar}}, \bibinfo {author} {\bibfnamefont {Jens-Christian}\ \bibnamefont {R{\" o}per}}, \bibinfo {author} {\bibfnamefont {Benoit}\ \bibnamefont {Aigouy}}, \bibinfo {author} {\bibfnamefont {Suzanne}\ \bibnamefont {Eaton}}, \ and\ \bibinfo {author} {\bibfnamefont {Frank}\ \bibnamefont {J{\" u}licher}},\ }\bibfield  {title} {\enquote {\bibinfo {title} {The influence of cell mechanics, cell-cell interactions, and proliferation on epithelial packing},}\ }\href {\doibase 10.1016/j.cub.2007.11.049} {\bibfield  {journal} {\bibinfo  {journal} {Current Biology}\ }\textbf {\bibinfo {volume} {17}},\ \bibinfo {pages} {2095–2104} (\bibinfo {year} {2007})}\BibitemShut {NoStop}%
\bibitem [{\citenamefont {Staddon}\ \emph {et~al.}(2023)\citenamefont {Staddon}, \citenamefont {Hernandez}, \citenamefont {Bowick}, \citenamefont {Moshe},\ and\ \citenamefont {Marchetti}}]{staddon_role_2023}%
  \BibitemOpen
  \bibfield  {author} {\bibinfo {author} {\bibfnamefont {Michael~F.}\ \bibnamefont {Staddon}}, \bibinfo {author} {\bibfnamefont {Arthur}\ \bibnamefont {Hernandez}}, \bibinfo {author} {\bibfnamefont {Mark~J.}\ \bibnamefont {Bowick}}, \bibinfo {author} {\bibfnamefont {Michael}\ \bibnamefont {Moshe}}, \ and\ \bibinfo {author} {\bibfnamefont {M.~Cristina}\ \bibnamefont {Marchetti}},\ }\bibfield  {title} {\enquote {\bibinfo {title} {The role of non-affine deformations in the elastic behavior of the cellular vertex model},}\ }\href {\doibase 10.1039/D2SM01580C} {\bibfield  {journal} {\bibinfo  {journal} {Soft Matter}\ }\textbf {\bibinfo {volume} {19}},\ \bibinfo {pages} {3080--3091} (\bibinfo {year} {2023})}\BibitemShut {NoStop}%
\end{thebibliography}%

\end{document}


\title{Supplemental Information of the manuscript "Cell divisions imprint long lasting elastic strain fields in epithelial tissues"}

\author{Ali Tahaei}
\affiliation{Max Planck Institute for the Physics of Complex Systems, 	N\"{o}thnitzer Str.38, 01187 Dresden, Germany}

\author{Romina Piscitello-G\'omez}
\affiliation{Max Planck Institute for Molecular Cell Biology and Genetics, Pfotenhauerstrasse 108, Dresden, 01307, Germany}

\author{S Suganthan}
\affiliation{Max Planck Institute for the Physics of Complex Systems, 	N\"{o}thnitzer Str.38, 01187 Dresden, Germany}

\author{Greta Cwikla}
\affiliation{Excellence Cluster, Physics of Life, Technische Universit\"{a}t Dresden, Arnoldstrasse 18, Dresden, 01307, Germany}

\author{Jana F. Fuhrmann}
\affiliation{Max Planck Institute for Molecular Cell Biology and Genetics, Pfotenhauerstrasse 108, Dresden, 01307, Germany}
\affiliation{Excellence Cluster, Physics of Life, Technische Universit\"{a}t Dresden, Arnoldstrasse 18, Dresden, 01307, Germany}

\author{Natalie A. Dye}
\affiliation{Max Planck Institute for Molecular Cell Biology and Genetics, Pfotenhauerstrasse 108, Dresden, 01307, Germany}
\affiliation{Excellence Cluster, Physics of Life, Technische Universit\"{a}t Dresden, Arnoldstrasse 18, Dresden, 01307, Germany}

\author{Marko Popovi\'c}
\affiliation{Max Planck Institute for the Physics of Complex Systems, 	N\"{o}thnitzer Str.38, 01187 Dresden, Germany}
\affiliation{Excellence Cluster, Physics of Life, Technische Universit\"{a}t Dresden, Arnoldstrasse 18, Dresden, 01307, Germany}
\affiliation{Center for Systems Biology Dresden, Pfotenhauerstrasse 108, 01307 Dresden, Germany}

\maketitle

\section{Continuum theory of a force dipole perturbation in two dimensions}
In this section, we derive the strain field generated by a force dipole located at the origin. The force dipole is fully described by the force dipole tensor $d_{ij}$.
First, we consider an elastic medium and we derive the steady-state strain field. Then, we consider the relaxation dynamics to the steady-state and compare viscoelastic and frictional relaxations as two sources of dissipation processes.

The strain field is given by $u_{ij}=(\partial_i u_j + \partial_j u_i)/2$, where $u_i$ is the displacement field. The strain field can be decomposed as
\begin{align}
    u_{ij} = \frac{1}{2}u_{kk} \delta_{ij} + \tilde{u}_{ij},
\end{align}
where $u_{kk}/2$ is the isotropic strain and $\tilde{u}_{ij}$ is the shear strain.
Similarly, for a symmetric force dipole, we write
\begin{align}
    D_{ij} = D_0 \delta_{ij} + \tilde{D}_{ij},
\end{align}
with $D_0$ the isotropic force dipole and $\tilde{D}_{ij}$ the traceless symmetric part of the force dipole. 
A schematic representation of strain field and force dipole decompositions is presented in Fig.1 (B.\textit{i}).

\subsection{Strain field of a force dipole: steady state}
We start by writing the force balance equation 
\begin{align}
    \partial_{j}\sigma_{ij} + f_i = 0,
    \label{eq:force_balance_ss}
\end{align}
where $\sigma_{ij}$ is the stress tensor and the force density of the force dipole is given by
\begin{align}
    f_i = \partial_{j}[\delta(\vec r) D_{ij}].
\end{align}
The constitutive linear elasticity equations are
\begin{align}
    P &= -\frac{\overline{K}}{2} u_{kk}, \label{eq:linear_elasticity_eqs1} \\
    \tilde{\sigma}_{ij} &= 2K \tilde{u}_{ij},
    \label{eq:linear_elasticity_eqs2}
\end{align}
where $P= -\sigma_{kk}/2$ is the two-dimensional pressure, $\tilde\sigma_{ij}$ is the traceless-symmetric component of the stress tensor and we denote it as shear stress, and $\overline{K}$ and $2K$ are bulk and shear moduli, respectively. Solving Eqs.~\ref{eq:linear_elasticity_eqs1} \& \ref{eq:linear_elasticity_eqs2} and Eq.~\ref{eq:force_balance_ss} in Fourier space, we find the strain field to be
\begin{align}
    u_{kk} &= -\frac{1}{K}\frac{\mu}{1+\mu}\frac{q_i q_j D_{ij}}{q^2} \\
    \tilde u_{ij} &= -\frac{1}{2K}\frac{D_{kl}q_l }{q^2} \left\lbrace q_i\delta_{jk} + q_j \delta_{ik} - \frac{q_k}{1+\mu} \left[ \mu \delta_{ij} + \frac{2q_i q_j}{q^2}\right] \right\rbrace,
\end{align}
where $q_i$ are the Fourier wavevectors, $q^2=q_x^2+q_y^2$ and $\mu=2K/\overline{K}$.
Finally, we apply the inverse Fourier transform to find the strain field in the real space, given by Eq.~2 in the main text. It is convenient to express the strain field using a Voigt-type tensor representation
\begin{align}
    \begin{pmatrix}
    u_{kk}(\vec{r}) \\ \tilde u_{xx}(\vec{r}) \\ \tilde u_{xy}(\vec{r}) 
    \end{pmatrix} = \frac{1}{2\pi K (1+\mu)}\mathbf{G}(\vec{r})
    \begin{pmatrix}
    D_{0} \\ \tilde D_{xx} \\ \tilde D_{xy} 
    \end{pmatrix},
    \label{eq:elastic_strain_dipole}
\end{align}
where
\begin{align}
    \mathbf{G}(\vec{r}) = \frac{1}{r^2}
    \begin{pmatrix}
    \mathcal{N} & -\mu{\cos 2\varphi} & -\mu{\sin 2\varphi} \\
    -\mu{\cos 2\varphi} & -{\cos 4\varphi} & -{\sin 4\varphi} \\
    -\mu{\sin 2\varphi} & -{\sin 4\varphi} & {\cos 4\varphi}
    \end{pmatrix},
    \label{eq:elastic_kernel}
\end{align}
and $\mathcal{N}$ is zero in an infinite system and a constant in a finite system. Elements of the propagator matrix $\mathbf{G}(\vec{r})$ are shown in Fig.~1 (B.\textit{iii}).
We define the normalized force dipole as
\begin{align}\label{eq:normalizedDipole}
    \bm{d} \equiv \frac{\bm{D}}{2K},
\end{align}
which will be the value that we obtain by fitting the linear elastic theory to experimental data. Namely, since our experimental measurements are purely geometric we can only determine force dipoles and stress fields normalized by an elastic constant.

\subsection{Strain field of a force dipole: relaxation}
We consider two sources of dissipation in an elastic sheet: viscosity and friction. For each of these cases, we study the strain field relaxation and we discuss how to distinguish these two different dissipation mechanisms from strain field relaxation dynamics.

\subsubsection{Frictional dissipation}\label{sec:frictional_relax}
We introduce a friction term $-\zeta \dot u_i$ as the source of dissipation in the force balance equation.
Then, from the force balance we have
\begin{align}
\label{eq:MomEq_fricton}
    -\partial_i P + \partial_j \tilde\sigma_{ji} + f_i - \zeta \dot u_i = 0.
\end{align}
After combining Eqs.~\ref{eq:linear_elasticity_eqs1} \& \ref{eq:linear_elasticity_eqs2}, and Eq.~\ref{eq:MomEq_fricton}, we find
\begin{align}
    \frac{\overline{K}}{2}\partial_i u_{kk} + 2K \partial_k \tilde{u}_{ik} + f_i - \zeta \dot{u}_i = 0,
\end{align}
and in the Fourier space we find the solutions
\begin{align}\label{eq:friciton_F_space_iso}
    u_{kk} &= -\frac{1}{K}\frac{\mu}{1+\mu}\frac{q_i q_j D_{ij}}{q^2}(1-e^{- q^2 t/\beta_1}), 
\end{align}
and
\begin{align}\begin{split}
    \tilde u_{ij} = &-\frac{1}{2K} \frac{q_m D_{lm}}{q^2} \times \\ &\Bigg\lbrace \left[ q_i \delta_{jl} + q_j \delta_{il} - 2 \frac{q_i q_j q_l}{q^2} \right](1 - e^{- q^2 t/\beta_2}) 
    \\
    &-\frac{\mu}{1+\mu}q_l\left[ \delta_{ij} - \frac{q_i q_j}{q^2} \right](1 - e^{-q^2 t / \beta_1}) \Bigg\rbrace \label{eq:friciton_F_space_traceless}.
\end{split}
\end{align}
Here, $\beta_1 = \frac{\zeta}{K}\frac{\mu}{1+\mu}$ and $\beta_2 = \frac{\zeta}{K}$.
The relaxation terms in Eqs.~\ref{eq:friciton_F_space_iso} \& \ref{eq:friciton_F_space_traceless} depend on $q$, so that length scale on which the system has relaxed will change in time i.e. the strain field propagates through space in time. To illustrate this, we plot the second angular mode of the strain field component $\tilde{u}_{xy}(r,m=2,t)$ in Fig.~2 (A.\textit{ii}), which clearly shows a propagating front. Beyond the front, the strain field rapidly vanishes and inside the front, it has relaxed to the steady state value. 

\subsubsection{Viscoelastic relaxation}\label{sec:viscous_relax}
We use a Kelvin-Voigt type of material equation to introduce viscous dissipation in the model of an elastic sheet. 
The constitutive equations are
\begin{align}
    P &= - \frac{\overline{K}}{2} u_{kk} - \frac{\overline{\eta}}{2} \dot{u}_{kk},
    \label{eq:linear_ViscoElasticity_eqs1}
    \\
    \tilde\sigma_{ij} &= 2K\tilde{u}_{ij} + 2\eta \dot{\tilde u}_{ij}.
    \label{eq:linear_ViscoElasticity_eqs2}
\end{align}
Here, $\overline{K}$ and $\overline{\eta}$ are bulk moduli and bulk viscosity, $2K$ and $2\eta$ are shear moduli and shear viscosity, and $\dot{\bm{u}}$ shows the temporal derivative of the strain field.
After combining Eqs.~\ref{eq:linear_ViscoElasticity_eqs1} \& \ref{eq:linear_ViscoElasticity_eqs2}, and Eq.~\ref{eq:force_balance_ss}, we find
\begin{align}
    \frac{\overline{K}}{2}\partial_i u_{kk} + \frac{\overline{\eta}}{2}\partial_i \dot{u}_{kk} + 2K \partial_k \tilde{u}_{ij} + 2\eta \partial_k \dot{\tilde{u}}_{ij} - f_i = 0.
    \label{eq:viscous_diff}
\end{align}
We solve Eq.~\ref{eq:viscous_diff} in Fourier space and find the transient isotropic and shear strain as,
\begin{align}
    u_{kk} &= -\frac{1}{K}\frac{\mu}{1+\mu}\frac{q_i q_j D_{ij}}{q^2}(1-e^{- t/\tau_1})
    \label{eq:iso_strain_Fourier_viscoelastic}
\end{align}
\begin{multline}
    \tilde{u}_{ij} = -\frac{1}{2K}\frac{D_{kl}q_l}{q^2} \Bigg\lbrace \left[ q_i\delta_{jk} + q_j \delta_{ik} - \frac{2q_i q_j q_k}{q^2} \right] (1 - e^{- t/\tau_2})
    \\
    - \frac{\mu}{1+\mu} q_k \left[ \delta_{ij} - \frac{2q_i q_j}{q^2} \right] (1-e^{- t/\tau_1}) \Bigg\rbrace ,
\label{eq:shear_strain_Fourier_viscoelastic}
\end{multline}
with two timescales $\tau_1 = \frac{\overline{\eta}}{\overline{K}}\frac{1+\mu^*}{1+\mu}$ and $\tau_2 = \eta / K$, where $\mu^*=2\eta/\overline{\eta}$. Unlike the case with friction, the relaxation factors do not depend on $q$ which means that there is no propagating front and the spatial profile of the strain field only changes in magnitude but not in shape.
After applying the inverse Fourier transformation to Eqs.~\ref{eq:iso_strain_Fourier_viscoelastic} \& \ref{eq:shear_strain_Fourier_viscoelastic}, and together with Eq.~\ref{eq:normalizedDipole} we find that,
\begin{align}
    \begin{pmatrix}
    u_{kk} \\ \tilde u_{xx} \\ \tilde u_{xy} 
    \end{pmatrix} = \frac{1}{2\pi K(1+\mu)}\mathbf{G}(\bm{r},t)
    \begin{pmatrix}
    D_{0} \\ \tilde D_{xx} \\ \tilde D_{xy} 
    \end{pmatrix}.
    \label{eq:viscoelastic_strain_dipole}
\end{align}
Here, $\mathbf{G}$ is the matrix whose components are given by $G_{ij}(\bm{r},t) = G_{ij}(\bm{r})T_{ij}(t)$, where no sum over repeated indices is performed. $G_{ij}(\bm{r})$ is the elastic kernel from Eq.~\ref{eq:elastic_kernel}, $T_{ij}(t)$ are components of matrix $\mathbf{T}$
\begin{align}
    \mathbf{T} =
    \begin{pmatrix}
    g_1(t) & g_1(t) & g_1(t) \\
    g_1(t) & (1+\mu)g_2(t) - \mu g_1(t) & (1+\mu)g_2(t) - \mu g_1(t) \\
    g_1(t) & (1+\mu)g_2(t) - \mu g_1(t) & (1+\mu)g_2(t) - \mu g_1(t)
    \end{pmatrix},
    \label{eq:viscous_temporal}
\end{align}
where,
\begin{align}
    g_1(t) &= 1-e^{-t / \tau_1},
    \label{eq:viscous_g1}\\
    g_2(t) &= 1-e^{-t/\tau_2}.
    \label{eq:viscous_g2}
\end{align}

Our results show that there is a qualitative difference in the dynamics of strain field relaxation between frictional and viscous dissipation mechanisms.
During the viscous relaxation of an elastic sheet, the radial dependence of the strain field is given by $1/r^2$ at all times, as shown in Eq.~\ref{eq:viscoelastic_strain_dipole}, whereas during frictional relaxation a propagation front appears. We show the radial profile of $\tilde{u}_{xy}(r,m=2, t)$ for viscous and frictional relaxation in Fig.~2 A. 
Therefore, the presence of a propagating front would be a signature of frictional dissipation. However, we do not find any sign of such a front in experimentally measured strain fields, which suggests that the dominant dissipation mechanism in the \textit{D. melanogaster} pupal wing epithelium is viscous on the timescales and lengthscales of the ablation experiments. 

Finally, to determine the viscous relaxation timescales we fit  Eq.~\ref{eq:viscoelastic_strain_dipole} to the measured transient strain field at each timepoint $t_i$. Since we have already determined the normalized force dipole tensor $\bm{d}$ and $\mu$, this allows us to find $g_1(t_i)$ and $g_2(t_i)$. We plot the obtained values of $1-g_1(t_i)$ and $1-g_2(t_i)$ in SI Fig.~\ref{fig:expData_viscousFit}. We then extract the viscous relaxation timescales $\tau_1$ and $\tau_2$ by fitting an exponential function to these data, see dashed lines in SI Fig.~\ref{fig:expData_viscousFit}.

\section{Experimental data}
\begin{figure}
    \centering
    \includegraphics[width=0.98\linewidth]{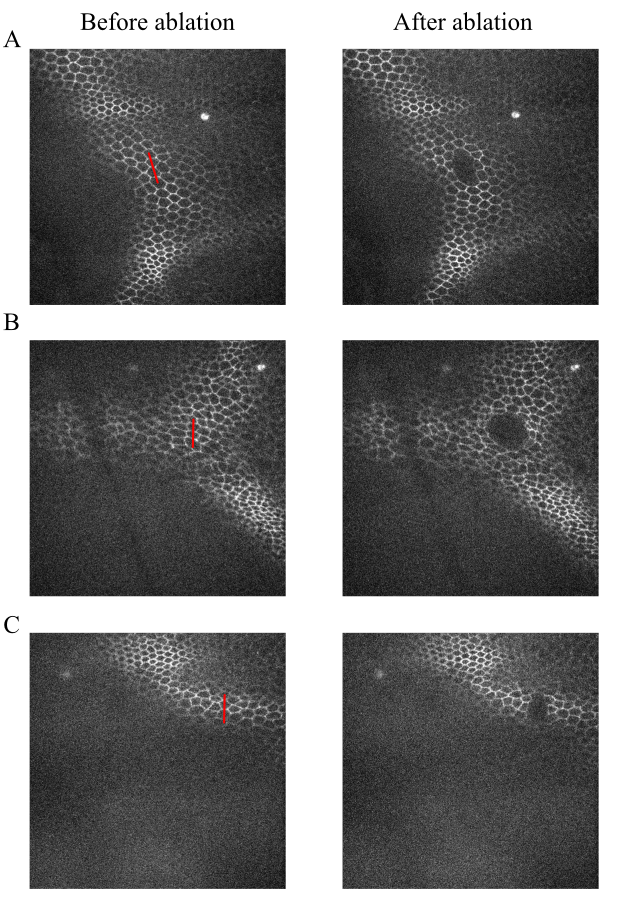}
    \caption{Snapshots of the apical surface showing fluorescently labeled cell outlines from three linear ablation experiments. The red line indicates the location of the laser ablation. Following the cut, the tissue deforms around the ablated region. However, quantifying this deformation is challenging due to large black areas near the cut, where the tissue is out of focus. These black regions highlight the experimental difficulty of maintaining proper imaging conditions in the immediate vicinity of the ablation.}\label{fig:poor_movies}
\end{figure}

As mentioned in the main text, the experiments presented in this manuscript were selected from the experiments reported in Ref.~\cite{Piscitello2022} based on our ability to segment and track cells in a radius of $40~\mu m$ around the laser ablation. In SI Fig.~\ref{fig:poor_movies}, we show examples of typical experiments that were not selected. In these experiments, tracking and segmenting the cell around the ablation is challenging due to large dark areas near the ablation, where the tissue is out of the imaging plane. 

Nevertheless, we were able to identify some experiments that allowed for the measuring of cell strain in a sufficiently large radius around the ablation. Furthermore, we observed that the tissue eventually initiates retrograde flows towards the ablation at different times in different experiments. We suspect this is the wound healing response of the tissue. In some cases, the wound healing process begins before the tissue reaches mechanical equilibrium or a static state. While the initial short-timescale response in these experiments can still be captured by linear viscoelastic theory (i.e., the Kelvin–Voigt model), the fitting process becomes more challenging. Here, we present one such experiment, referred to as Experiment 2, see below for more details. The experiment presented throughout the main text is referred to as Experiment 1. In addition, we include two further experiments, Experiments 3 and 4, which appear to reach mechanical equilibrium but could only be segmented over a smaller spatial range compared to Experiment 1. Additional information about each of the experiments is presented in Table 1. 

For Experiment 2, the strain field at $50$ seconds after the ablation in SI Fig. \ref{fig:sup_strainField_expData2} A shows a clear angular pattern around the cut. The angular Fourier modes averaged over the time interval between $50 - 90$ seconds after the ablation are shown in SI Fig.~\ref{fig:sup_strainField_expData2} B. A clear signal in the $m=2$ and $m=4$ modes suggests the angular pattern is consistent with the predictions of linear elastic theory. We find that the radial profile of the strain field is also consistent with that of the linear elastic sheet on short timescales $\approx 10$ second equal to the relaxation timescales $\tau_1$ and $\tau_2$ measured in Experiment 1, see SI Fig.~\ref{fig:sup_transient_ablation_expData} and SI Fig.~\ref{fig:sup_transient_ablation_expData2} for comparison. At longer times, we compare the radial strain profiles of Experiments 1 and 2 in SI Fig.~\ref{fig:sup_transient_ablation_LongTime}, and observe deviations in Experiment 2 that suggest additional effects, possibly due to the influence of wound healing on the decay of the strain field. 

In SI Fig. \ref{fig:sup_strainField_prediction}, we show the fits of the linear elastic theory to the observed strain field pattern of all of the experiments over a range of $12~\mu m$ to $25~\mu m$ from the center of the cut. For Experiments 1, 3, and 4 (panels A, B, and C), the fits are performed during time intervals in which the system appears to have reached a mechanical equilibrium. As discussed, Experiment 2 does not reach a mechanical equilibrium before the onset of wound healing. Therefore, the fit for Experiment 2 is performed over the interval from $50$ to $90$ seconds, which was selected as the time window most closely approximating a "quasi-static" state. All the fitting parameters are summarized in Table~\ref{tab:my_label}. We find that the ratio of shear to bulk modulus $\mu$ remains comparable across all four experiments. In contrast, the magnitude of the inferred force dipoles varies between experiments, suggesting that pre-stress levels differ across cells.

\begin{table*}[!ht]
    \centering
    \begin{tabular}{|c|ccccccc|}
        \hline
         & Stage & $d_0 ~ [\mu m ^2]$ & $\tilde{d}_{xx} ~ [\mu m ^2]$ & $\tilde{d}_{xy} ~ [\mu m ^2]$ & $\mu$ & $\tau_1 ~ [\rm{seconds}]$ & $\tau_2 ~ [\rm{seconds}]$ \\ [1ex]\hline\hline
         Exp. 1 & $20$ hAPF  & $160 \pm 10$ & $150 \pm 10$ & $20 \pm 10$  & $1.9 \pm 0.1$ & $11 \pm 1$ & $12 \pm 1$ \\[1ex]
         Exp. 2 & $20$ hAPF  & $ 150\pm 20$ & $310 \pm 20$  & $0 \pm 20$ & $2.1 \pm 0.3$ & $48 \pm 2$ & $55 \pm 2$ \\[1ex]
         Exp. 3 & $20$ hAPF  & $ 120\pm 10 $ & $ 44\pm 4 $ & $ 33\pm 9$  & $2.5 \pm 0.6$ & $8.5 \pm 0.5$ & $8.0 \pm 0.5$ \\[1ex]
         Exp. 4 & $24$ hAPF  & $ 80 \pm 10$ & $99 \pm 7$ & $-30\pm 9$ & $2.5 \pm 0.2$ & $27 \pm 2$ & $24 \pm 2$ \\[1ex] \hline
    \end{tabular}
    \caption{Mechanical parameters obtained by fitting the Kelvin-Voigt visco-elastic theory to the observed strain field pattern in four different experiments. The data from Experiment 1 is presented in the main text.}\label{tab:my_label}
\end{table*}

Finally, we compare the time evolution of the radial profiles of the strain field's angular Fourier modes $m=2$ of Experiments 2, 3, and 4 in SI Fig.~\ref{fig:sup_transient_ablation_RestExp}. We find that at early times the radial profiles are consistent with the $r^{-2}$ decay, consistent with a Kelvin-Voigt viscoelastic model, but then evolve additional features at longer times. We apply a fit of the viscoelastic model to find the viscous relaxation timescales $\tau_1$ and $\tau_2$. In SI Fig.~\ref{fig:expData_viscousFit}, the factors of the relaxation matrix elements $T_{ij}(t)$ are plotted against time and the fitted parameters are presented in Table \ref{tab:my_label}. In Experiments 1 and 3, the relaxation timescales are comparable and approximately $10$ seconds. In contrast, Experiment 2 exhibits a longer relaxation timescale of approximately $60$ seconds, coinciding with the onset of the additional movements that we interpret as the wound healing process.

\section{Vertex model implementation}\label{sec:VM_section}
To simulate linear ablation, we use the 2D vertex model of epithelial tissues \cite{Farhadifar2007}. The tissue is modeled as a tiling of polygonal cells and is endowed with the energy functional 

\begin{equation}
    W = \sum_{c \in \text{cells}} \frac{K_c}{2}(A_c-A^0_c)^2 + \sum_{b \in \text{bonds}} \Lambda_b L_b + \sum_{c \in \text{cells}} \frac{\Gamma_c}{2} P_c^2 
\end{equation}

In the above expression, the first term represents the area elasticity of cells arising from the volume incompressibility of the epithelial layer; $A_c$, $A^0_c$ and $K_c$ denote the area, preferred area and area elastic modulus of cell 'c', respectively. The second term corresponds to interfacial tension in the bonds arising from actomyosin contractility and adhesion between cells sharing the bond; $\Lambda_b$ and $L_b$ correspond to the bond tension parameter and the length of the bond 'b', respectively. The final term is quadratic in length resulting in a bond tension that is dependent on the perimeter of the cell; $\Gamma_c$ sets the strength of the perimeter contractility and $P_c$ is the perimeter of the cell. The degrees of freedom of the system are the vertex positions $\vec{r}_v$ which are evolved using overdamped dynamics as follows

\begin{equation}
    \gamma \frac{d\vec{r}_v}{dt} = -\nabla_{\vec{r}_v} W 
\end{equation}

We first prepare the sample by randomizing a hexagonal network of 2500 cells using bond tension fluctuations under periodic boundary conditions. We choose the parameters $\Lambda_b=\Lambda=0$, $\Gamma_c=\Gamma=0.11$, and $K_c=K= 1$, such that it corresponds to a negative Poisson's ratio as reported in \cite{staddon_role_2023} for a hexagonal lattice. Similar values of the Poisson ratio have also been measured in the \textit{D. melanogaster} pupal wing epithelium \cite{Piscitello2022}. We then apply pure shear stress to the sample to incorporate the presence of pre-stresses in the wing disc. Once the target value of the pure shear stress is reached we fix the simulation box boundaries and relax the system to the nearest energy minimum. 
We perform a linear ablation in the simulations in two steps; first, we remove a single bond from the cellular network by removing its vertices, and the two cells that originally abutted the bond are connected to form a single cell.
Then, for the new cell, we set   $K_c = \Gamma_c = 0$. The removal of the bond introduces a force dipole in the tissue, resulting in a redistribution of stress. During this process, we record the configuration of the system and perform the same analysis as done for the experimental data.

\subsection{Shear and bulk moduli of the vertex model}
\begin{figure}
    \centering
    \includegraphics[width=0.49\linewidth]{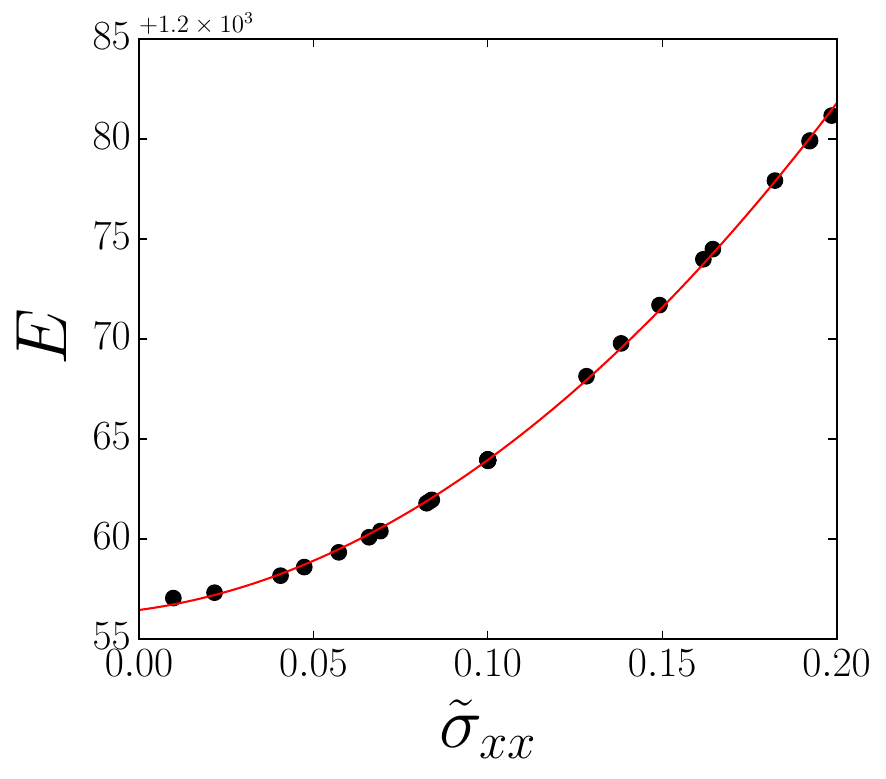}
    \includegraphics[width=0.49\linewidth]{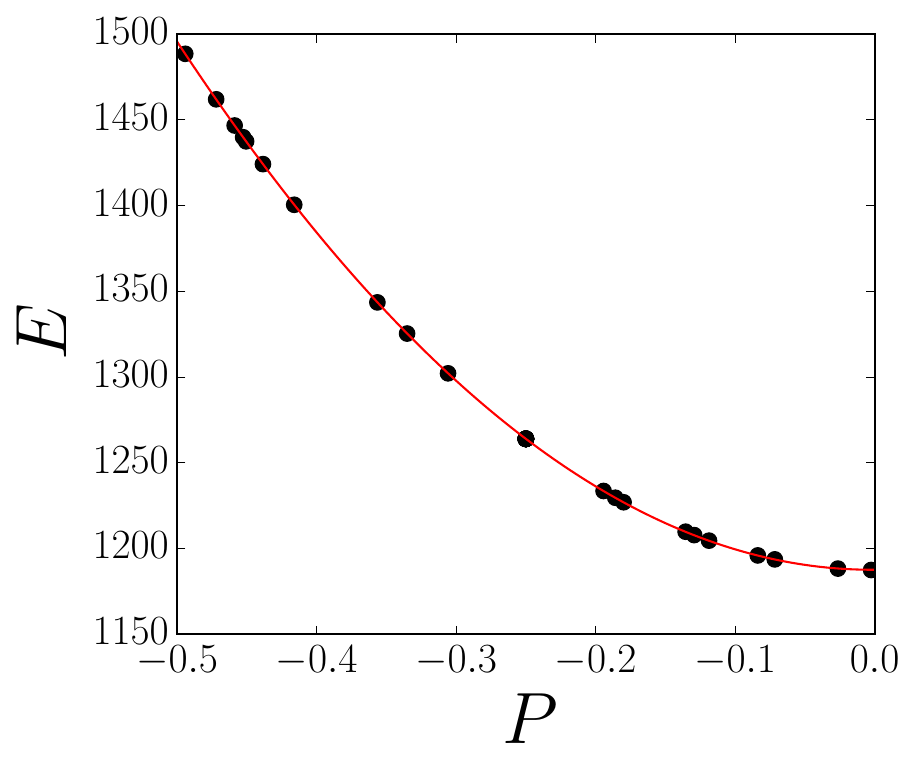}
    \caption{The energy of the vertex model against imposed (left) shear (B) pressure. Each data point is the energy after full relaxation of the network, and the red line presents the best fit around $\tilde{\sigma}_{xx}=0.1$ and $P=-0.25$.}
    \label{fig:periodicLoading_VM}
\end{figure}

To measure the Poisson ratio of the tissue independently from the method of elastic strains, we quantify the elastic strain by directly minimizing the energy of the vertex model exposed to varying shear stress values. 
We start from the initial network used for the ablation simulations with $\Lambda=0$, $\Gamma=0.11$, $P=-0.25$ and $\tilde{\sigma}_{xx}=0.1$. Then, we impose additional shear/isotropic stress on the system while isotropic/shear stress is kept constant and let the system relax. Fig.~\ref{fig:periodicLoading_VM} shows the resulting energy vs. shear/isotropic stress plots after the relaxation, with the red line showing the best fit describing the shear modulus $2K=(9.7 \pm 1.0) \times 10^{-4}$ and bulk modulus $\overline{K}=(8.1\pm0.8) \times 10^{-4}$, corresponding to a Poisson ratio of $\nu=-0.095 \pm 0.009$, consistent with $\nu=-0.13 \pm 0.04$ as we found through the ablation simulations.

\bibliography{bib.bib}

\clearpage

\begin{figure}
    \includegraphics[width=\linewidth]{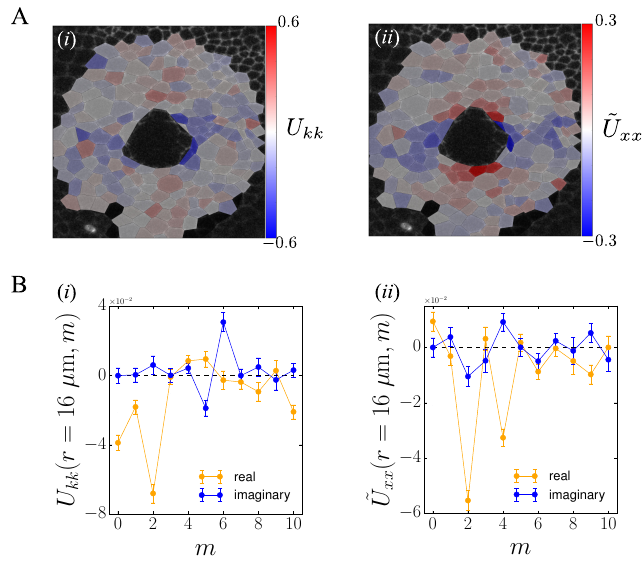}
    \caption{(A) Components $kk$ and $xx$ of the strain field generated by linear laser ablation in Experiment 1. Corresponding analysis of the $xy$ component is shown and described in the main text Fig. 1 A. (B) Angular Fourier modes of the strain field from Experiment 1 show the strongest signal in modes $m=0$ and $m= 2$ for $U_{kk}$ strain component, and modes $m=2$ and $m=4$ for $U_{xx}$ component. Real and imaginary parts of Fourier modes correspond to $\cos{(m \varphi)}$ and $\sin{(m \varphi)}$ angular profiles, respectively.}
    \label{fig:sup_strainField_expData}
\end{figure}

\begin{figure*}
    \includegraphics[width=.8\linewidth]{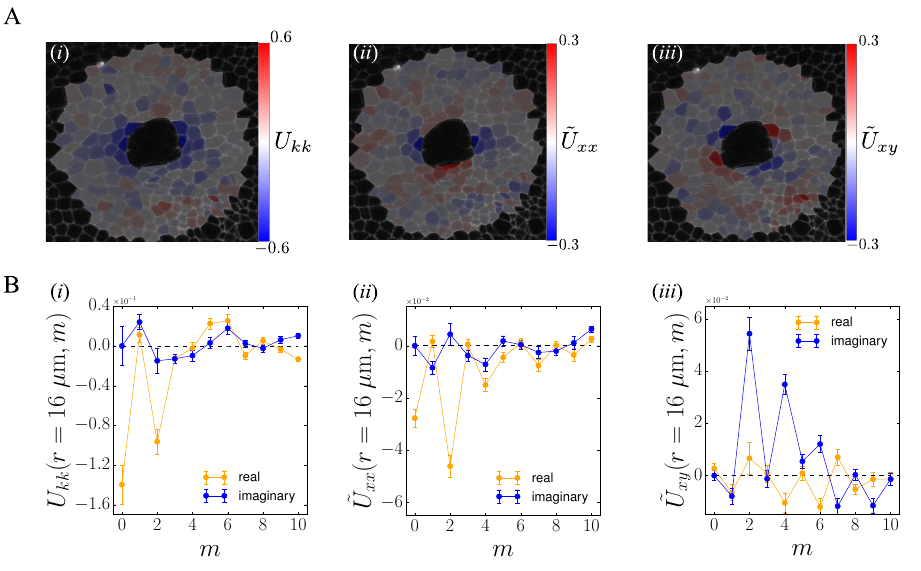}
    \caption{(A) All components of the strain field generated by linear laser ablation in Experiment 2. (B) Angular Fourier modes of the strain field from Experiment 2 show the strongest signal in modes $m=0$ and $m= 2$ for $U_{kk}$ strain component, and modes $m=2$ and $m=4$ for $U_{xx}$ and $U_{xy}$ components. Real and imaginary parts of Fourier modes correspond to $\cos{(m \varphi)}$ and $\sin{(m \varphi)}$ angular profiles, respectively. The analysis is performed over the data accumulated between $50 - 90$ seconds after the laser ablation.}
    \label{fig:sup_strainField_expData2}
\end{figure*}

\begin{figure}
    \centering
    \includegraphics[width=\linewidth]{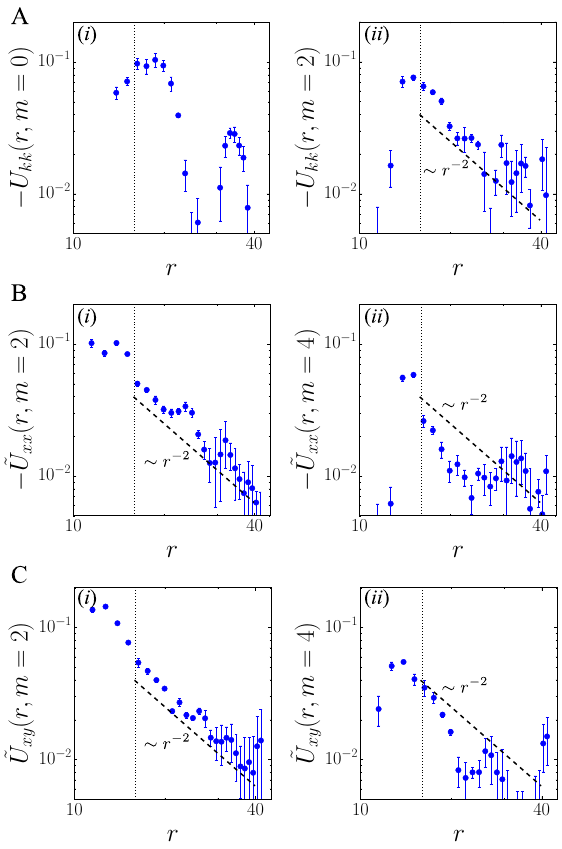}
    \caption{Radial dependence of strain field components angular Fourier modes, measured in Experiment 1. A vertical dotted line indicates $r= 16\mu m$. The dashed line indicates $r^{-2}$ decay for comparison, where appropriate. (A) Modes $m=0$ and $m=2$ of the strain component $kk$. (B) Modes $m=0$ and $m=2$ of the strain component $xx$. (C) Modes $m=0$ and $m=2$ of the strain component $kk$. The analysis is performed over the data accumulated between $30 - 70$ seconds after the laser ablation.}
    \label{fig:sup_strainField_expData}
\end{figure}

\begin{figure*}
    \centering
    \includegraphics[width=\linewidth]{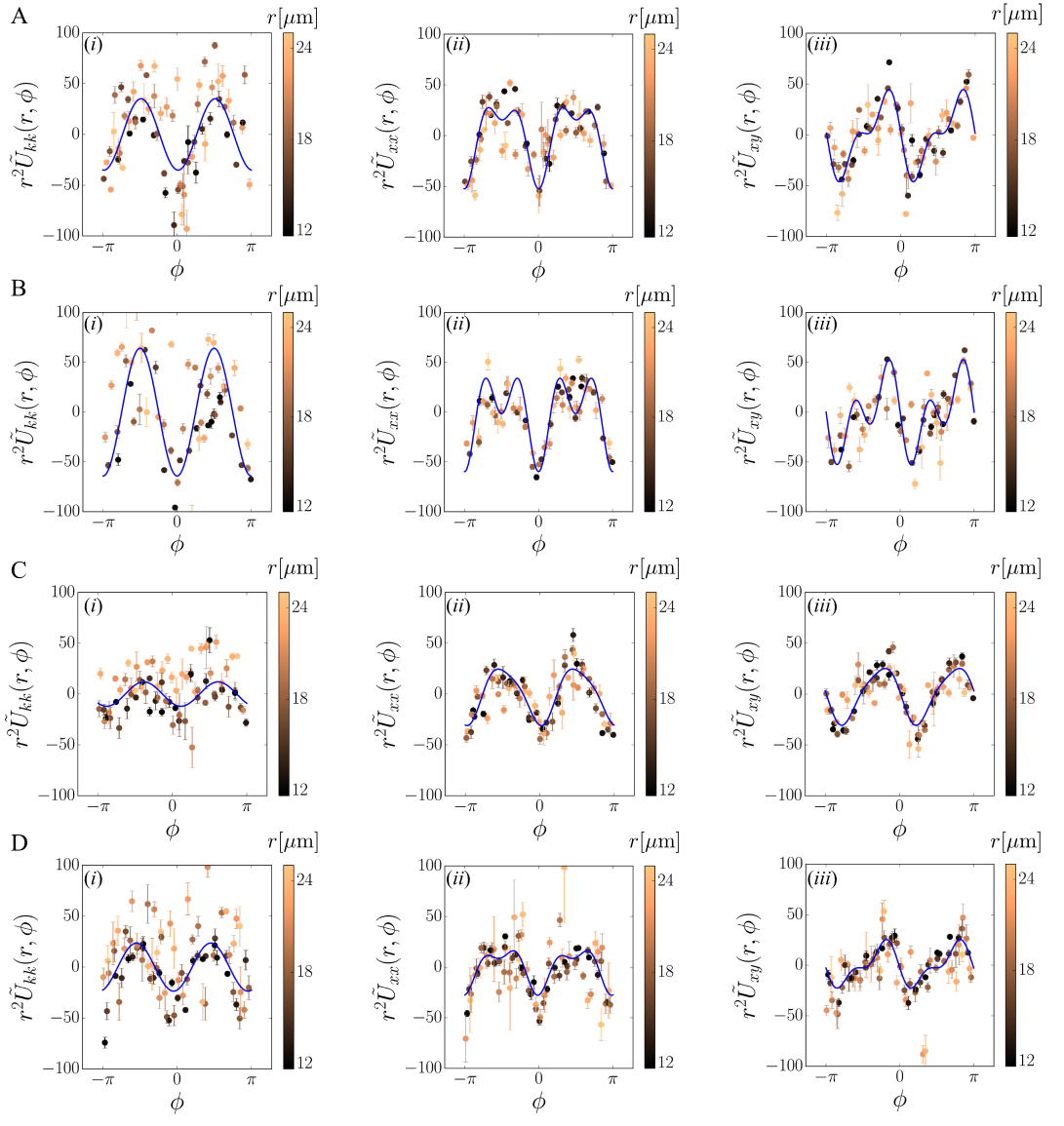}
    \caption{Blue line shows the fit of the linear elastic theory to the data points representing the strain field measured in laser ablation experiments, rows correspond to different experiments with  (A) Experiment 1, (B) Experiment 2, (C) Experiment 3, and (D) Experiment 4.  Data at different radial distances from the ablation are shown in different colors as indicated by the colorbar. Error bars indicate the standard deviation of measured strain values in the corresponding radial/angular bin.}
    \label{fig:sup_strainField_prediction}
\end{figure*}

\begin{figure}
    \centering
    \includegraphics[width=\linewidth]{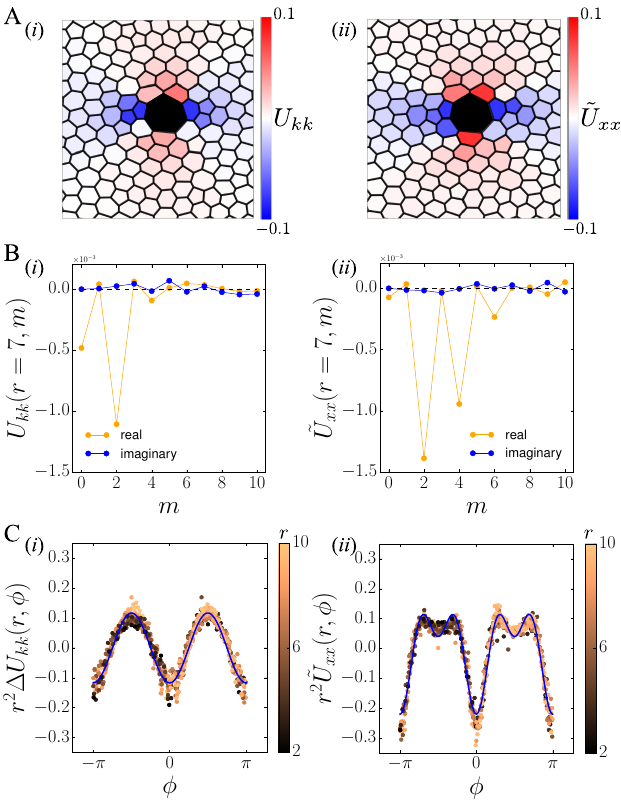}
    \caption{(A) Steady-state strain field components $U_{kk}$ and $\tilde{U}_{xx}$ generated by a laser ablation simulation in the vertex model. (B)  Angular Fourier modes of the strain field components show the strongest signal in modes $m=0$ and $m= 2$ for $U_{kk}$ strain component, and modes $m=2$ and $m=4$ for $U_{xx}$ component. Real and imaginary parts of Fourier modes correspond to $\cos{(m \varphi)}$ and $\sin{(m \varphi)}$ angular profiles, respectively. (C) The blue line shows the fit of the linear elastic theory to the data points representing the strain field measured in laser ablation experiments. Data at different radial distances from the ablation are shown in different colors as indicated by the colorbar. Data points correspond to strain measured in individual cells in the vertex model.}
    \label{fig:sup_strainField_VM}
\end{figure}

\begin{figure*}[h]
    \begin{minipage}[t]{0.48\textwidth}
        \centering
        \includegraphics[width=\textwidth]{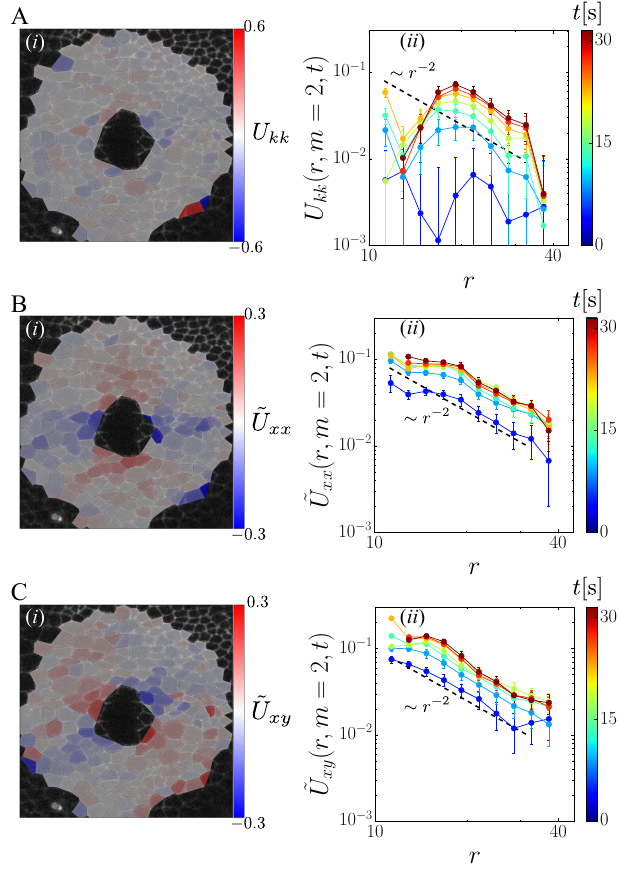}
        \caption{Time evolution of the strain field induced by the laser ablation in Experiment 1. Each panel shows a component of the strain field after relaxation on the left and radial dependence of the second Fourier mode on the right. (A) $U_{kk}$ component, (B) $\tilde{U}_{xx}$ component and (C) $\tilde{U}_{xy}$ component. The dashed line shows a power-law decay $r^{-2}$ for reference.}
        \label{fig:sup_transient_ablation_expData}
    \end{minipage}
    \hfill
    \begin{minipage}[t]{0.48\textwidth}
        \centering
        \includegraphics[width=\textwidth]{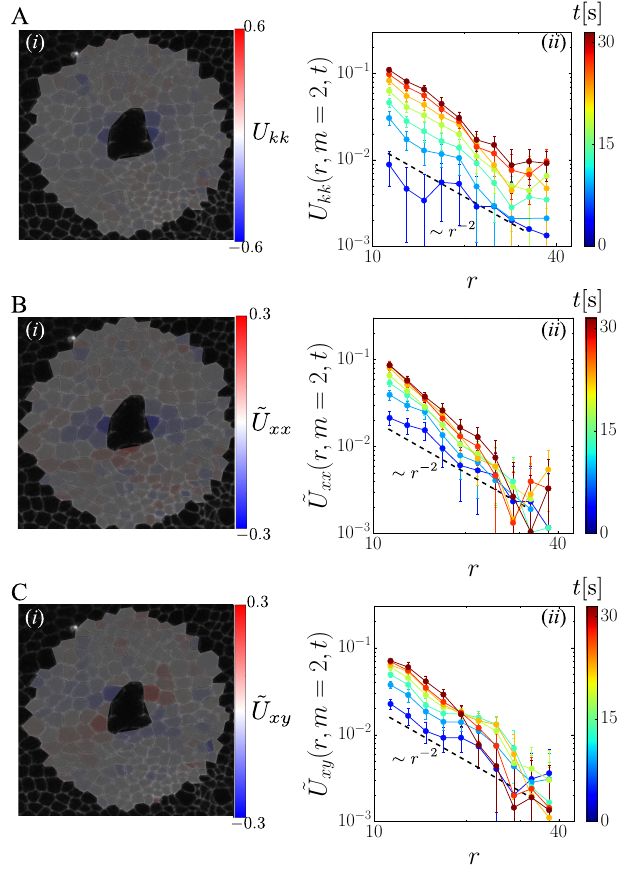}
        \caption{Time evolution of the strain field induced by the laser ablation in Experiment 2. Each panel shows a component of the strain field after relaxation on the left and radial dependence of the second Fourier mode on the right. (A) $U_{kk}$ component, (B) $\tilde{U}_{xx}$ component and (C) $\tilde{U}_{xy}$ component. The dashed line shows a power-law decay $r^{-2}$ for reference.}
        \label{fig:sup_transient_ablation_expData2}
    \end{minipage}
\end{figure*}

\begin{figure}
    \centering
    \includegraphics[width=.99\linewidth]{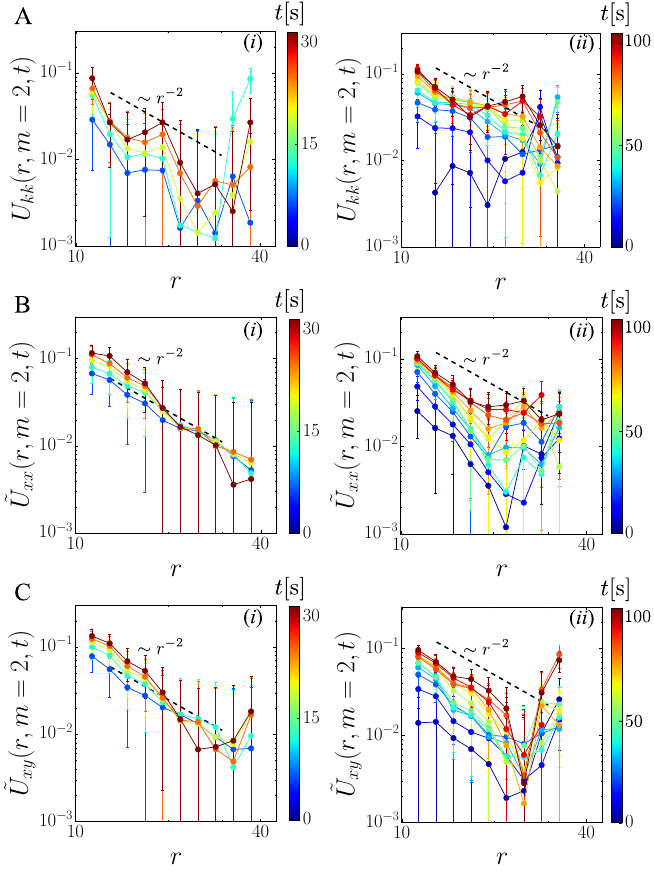}
    \caption{Time evolution of the radial profile of the strain field induced by a laser ablation over Experiments 3 and 4. Each row shows one component of the strain field (A) $U_{kk}$, (B) $\tilde{U}_{xx}$, and (C) $\tilde{U}_{xy}$. The columns show (\textit{i}) Experiment 3 and (\textit{ii}) Experiment 4. The black dashed line shows $1/r^2$ as a reference in all plots. }
    \label{fig:sup_transient_ablation_RestExp}
\end{figure}

\begin{figure}
    \centering
    \includegraphics[width=.98\linewidth]{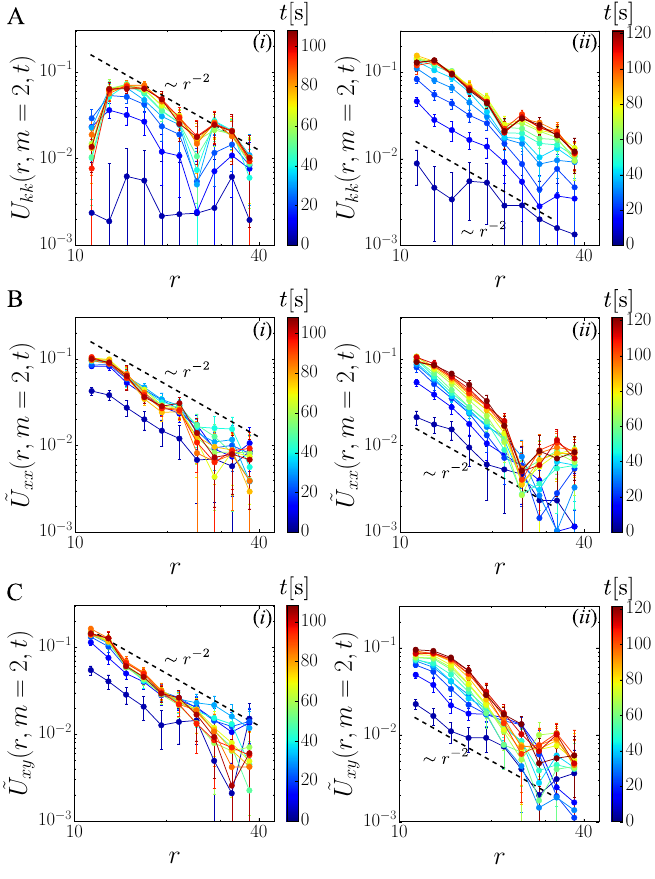}
    \caption{Comparison of the time evolution of the strain field induced by the laser ablation between Experiment 1 (\textit{i}) and Experiment 2 (\textit{ii}) at long times up to $120 s$ after ablation. A decay of the radial profile faster than $1/r^2$ is observed at longer times, probably due to the wound healing process. This effect is more extreme and appears earlier in Experiment 2. Each panel shows a component of the strain field after relaxation on the left and radial dependence of the second Fourier mode on the right. (A) $U_{kk}$ component, (B) $\tilde{U}_{xx}$ component and (C) $\tilde{U}_{xy}$ component. The dashed line shows a power-law decay $r^{-2}$ for reference.}
    \label{fig:sup_transient_ablation_LongTime}
\end{figure}

\begin{figure}
    \centering
    \includegraphics[width=\linewidth]{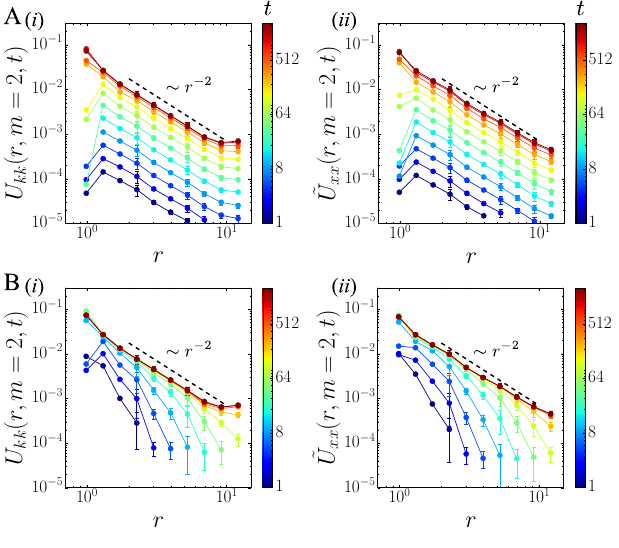}
    \caption{Time evolution of the second Fourier mode $m=2$ of the strain field components $kk$ and $xx$, induced by laser ablation in vertex model simulation with (A) viscous dissipation and (B) frictional dissipation. The dashed line indicates the power-law $r^{-2}$ decay for reference.}
    \label{fig:sup_transient_ablation_VM}
\end{figure}

\begin{figure}
    \centering
    \includegraphics[width=\linewidth]{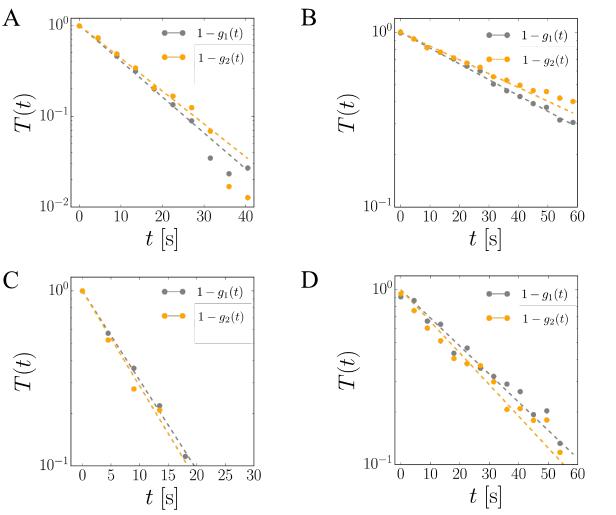}
    \caption{Factors of the relaxation matrix elements $T_{ij}(t)$ evaluated over time from the fit of the theory to the data. Here, $T(t)$ stands for the two factors $1 - g_1(t)$ and $1- g_2(t)$, as indicated by the legend. The dashed lines show the fits with the parameters reported in Table.~\ref{tab:my_label}. (A) The data from the laser ablation Experiment 1, presented in the main text. (B) The data from Experiment 2 are described here in the SI. In this experiment, the tissue does not reach a mechanical equilibrium state. (C) Experiment 3 and (D) Experiment 4.}
    \label{fig:expData_viscousFit}
\end{figure}

\begin{figure}
    \centering
    \includegraphics[width=\linewidth]{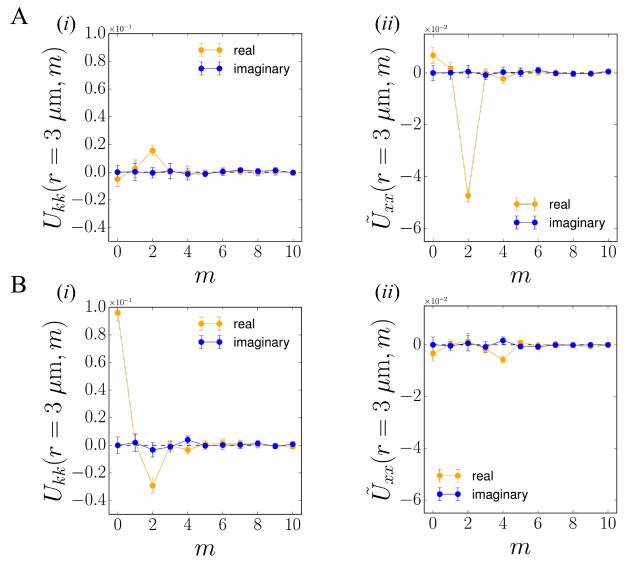}
    \caption{The Fourier modes of strain field due to cell divisions in wing disc at two different time points and $r=3 ~\mu m$. (A) At the peak of inflation, with $t=-20$ minutes. (B) At the time point with approximately the strongest anisotropic force dipole, that is $t=30$ minutes. }
    \label{fig:Divisions_AngularPattern}
\end{figure}

\begin{figure*}
    \centering
    \includegraphics[width=.75\textwidth]{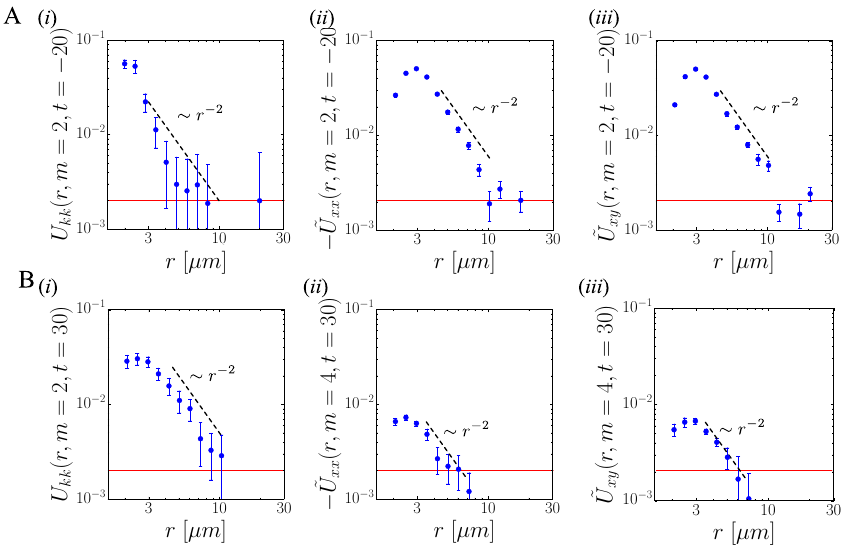}
    \caption{The radial profile of nonvanishing Fourier modes of strain field due to cell divisions in the wing disc at two different time points. (A) At the peak of inflation, with $t=-20$ minutes. (B) At the time point with approximately the strongest anisotropic force dipole, that is $t=30$ minutes. The black dashed line shows a radial decay of $1/r^2$. The radius of one cell is around $2 ~\mu m$, and the radius of the inflated mother cell at $t=-20$ minutes is around $4 ~\mu m$.}
    \label{fig:Divisions_radialProfile}
\end{figure*}

\begin{figure}
    \centering
    \includegraphics[width=.49\linewidth]{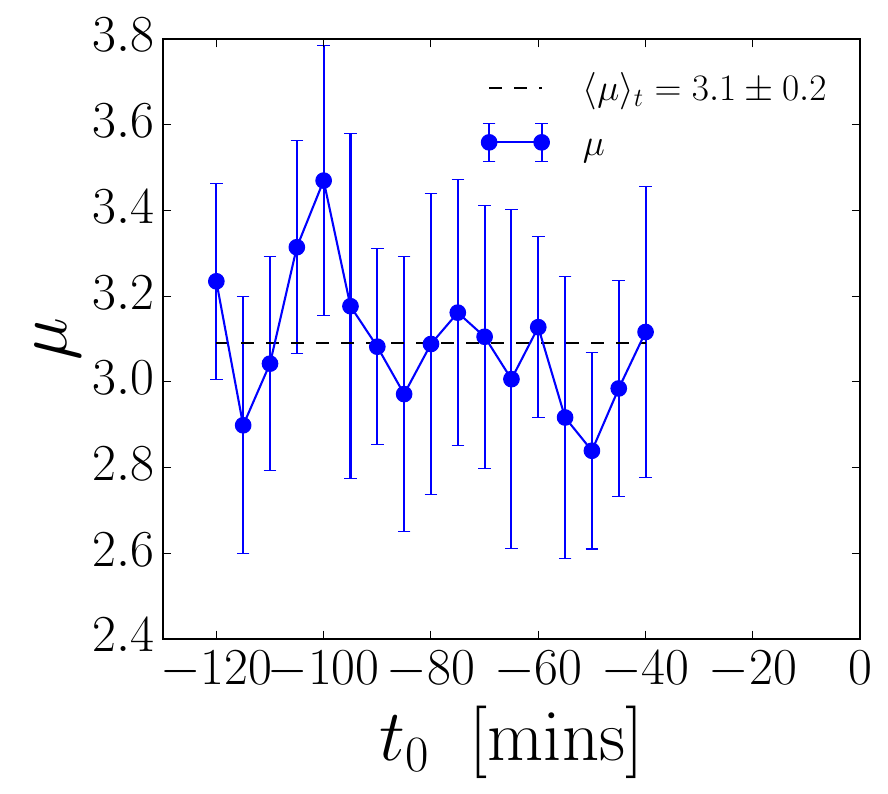}
    \caption{Fitting of $\mu$ by fixing $t=30~ \rm{minutes}$ and varying $t_0$. The dashed line shows the average over all the shown data points and the uncertainty interval is the standard deviation over this range.}
    \label{fig:mu_t0_plot}
\end{figure}

\begin{figure}
    \centering
    \includegraphics[width=\linewidth]{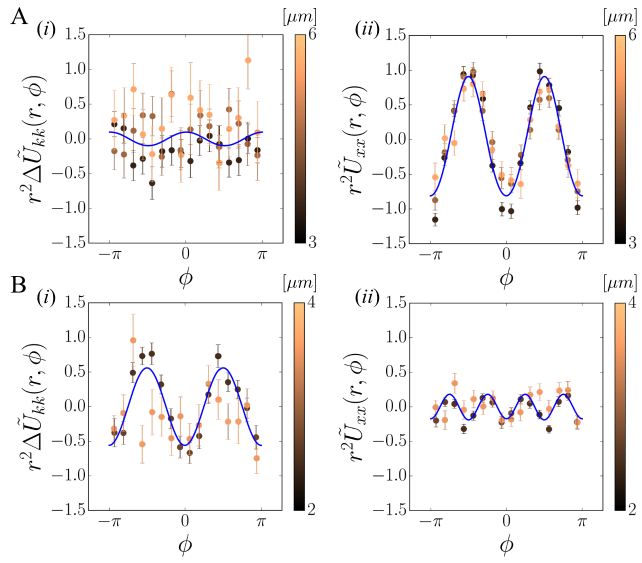}
    \caption{The comparison of data and fit. The data points are averages in polar bins. (A) At the peak of inflation with $t=-20$ minutes. (B) At the time point with approximately the strongest anisotropic force dipole, that is $t=30$ minutes. In all plots, the blue line shows the prediction. The error bars show the standard error measured over each bin.}
    \label{fig:data_fit_comp}
\end{figure}

\begin{figure}
    \centering
    \includegraphics[width=\linewidth]{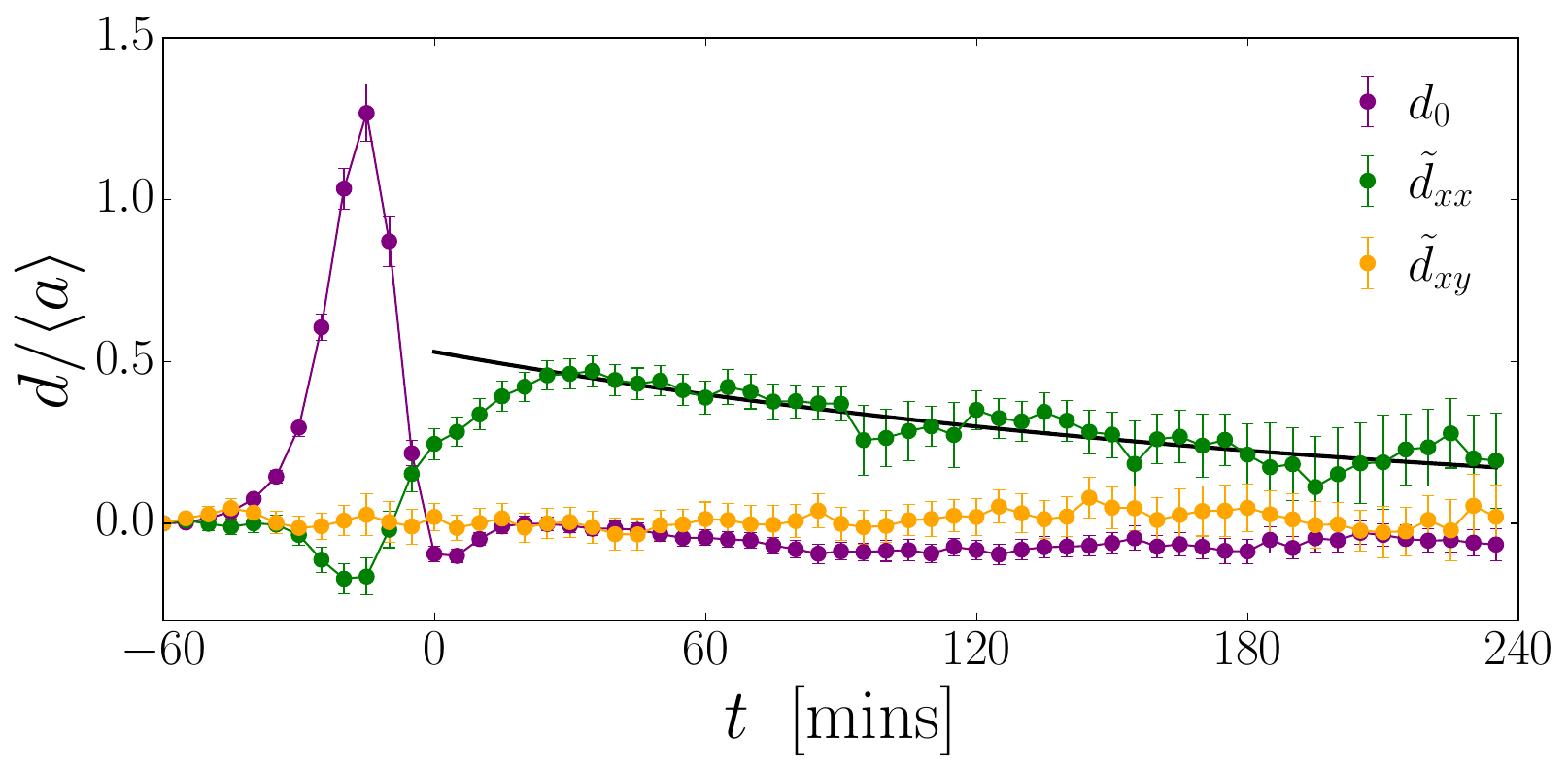}
    \caption{The force dipole generated by cell divisions at several time points as shown in Fig.~3 D. The black line shows the best fit to the anisotropic force dipole (green) with a timescale $\tau_\alpha=3.5 \pm 0.7$ hours and $\tilde{d}_{xx}/\langle a \rangle = 0.53 \pm 0.01$ as described in the methods.}
    \label{fig:tau_alpha_fit}
\end{figure}